\documentclass{pazhastlm}

\usepackage{graphicx}
\usepackage{natbib}

\usepackage{color}

\begin{document}

\journalinfo{2024}{50}{8}{1}[21]

\title{Humps on the profiles of the radial-velocity distribution and the age of the Galactic bar}

\author{A.~M. Melnik\address{1}\email{anna@sai.msu.ru},
  E.~N.~Podzolkova\address{1,2},
  \addresstext{1}{Sternberg Astronomical Institute, Lomonosov Moscow
State University, Universitetskij pr. 13, Moscow 119991,  Russia}
  \addresstext{2}{Faculty of Physics, Lomonosov Moscow State University, Leninskie
Gory 1-2, Moscow 119991, Russia} }

\shortauthor{Melnik and Podzolkova}

\shorttitle{Humps on the $V_R$-velocity profiles}

\submitted{September 2, 2024; in final form, September 12, 2024}

\begin{abstract}
We studied the model of the Galaxy with a bar which reproduces well
the distributions of the observed radial, $V_R$, and azimuthal,
$V_T$, velocities derived from the {\it Gaia} DR3 data along the
Galactocentric distance $R$. The model profiles of the distributions
of the velocity $V_R$ demonstrate a periodic increase and the
formation of a hump (elevation) in the distance range of 6--7 kpc.
The average amplitude and period of variations in the velocity $V_R$
are $A=1.76\pm0.15$ km s$^{-1}$ and $P=2.1\pm0.1$ Gyr. We calculated
angles $\theta_{01}$, $\theta_{02}$ and $\theta_{03}$ which determine
orientations of orbits relative to the major axis of the bar at the
time intervals: 0--1, 1--2 and 2--3 Gyr from the start of simulation.
Stars whose orbits change orientations as follows:
$0<\theta_{01}<45^\circ$, $-45<\theta_{02}<0^\circ$ and
$0<\theta_{03}<45^\circ$, make a significant contribution to the hump
formation.  The fraction of orbits trapped into libration among
orbits lying both inside and outside the Outer Lindblad Resonance
(OLR) is 28\%. The median period $P$ of long-term variations in the
angular momentum and total energy of stars increases as  Jacobi
energy approaches the values typical for the OLR but then sharply
drops. The distribution of model stars over the period $P$ has two
maxima located at $P=0.6$ and $1.9$ Gyr. Stars with orbits lying both
inside and outside the corotation radius (CR) concentrate to the
first maximum. The distribution of stars whose orbits lie both inside
and outside the OLR depends on their orientation: orbits elongated
perpendicular to the bar concentrate to the first maximum but those
stretched parallel to the bar concentrate to the second maximum. The
fact that the observed profile of the $V_R$-velocity distribution
derived from the {\it Gaia} DR3 data does not show a hump suggests
that the age of the Galactic bar, counted from the moment of reaching
its full power, must lie near one of two values: $2.0\pm0.3$ or
$4.0\pm0.5$ Gyr.

\end{abstract}

\section{1. Introduction}

Infrared observations provide irrefutable evidence that the Galaxy
includes a bar. The position angle of the bar relative to the Sun
amounts to $\sim 40^\circ$ \citep{dwek1995, benjamin2005,
cabrera-lavers2007, gonzalez2012, nesslang2016}. Estimates of the
angular velocity of the bar rotation,  $\Omega_b$, obtained, among
other things, from kinematical data, give a value of $\Omega_b$  in
the range of 40--60 km s$^{-1}$ kpc$^{-1}$ \citep[][and other
papers]{kalnajs1991, dehnen2000, minchev2007, gerhard2011,
antoja2014, monari2017, melnik2019,melnik2021, asano2022}.

The positions of the resonances of the  bar in the Galactic disk are
determined from the following condition:

\begin{equation}
\frac{m}{n}= \frac{\kappa}{\Omega-\Omega_b}, \label{resonance}
\end{equation}

\noindent where $m$ is the number of full epicyclic oscillations that
a star makes during $n$ revolutions relative to the bar.  The ratio
$m/n=2/1$ corresponds to the position of  the  Inner Lindblad
Resonance (ILR) and the ratio $m/n=-2/1$ determines the radius of the
Outer Lindblad Resonance (OLR). The ultraharmonic resonances
$m=\pm$4/1 are also important \citep{contopoulos1983,
contopoulos1989, athanassoula1992a}.

Estimates of the age of the Galactic bar differ significantly.
\citet{cole2002} studied the distribution of carbon stars in the
Galaxy and estimated the age of the Galactic bar to be less than 3
Gyr. The studies of  stellar ages in the Galactic bulge showed the
presence of a noticeable fraction of stars of intermediate ages (2--5
Gyr) \citep[][and other papers]{nataf2016, bensby2017,
hasselquist2020}.  The  resent  episodes of  star formation in the
bar region can be triggered by the formation of a strong bar.
\citet{nepal2024} studied a sample of super-metal-rich stars in the
solar neighborhood and found a sharp change in the age-metallicity
relation  $\sim 3$ Gyr ago which can indicate the end of the bar
formation era. \citet{bovy2019} estimated the age of the Galactic bar
to be $\sim8$ Gyr. \citet{debattista2019} modelled the X-shape for
stars of varying ages and found the age of the Galactic bar to be
greater than 5 Gyr. \citet{sanders2024} studied Mira variables in the
nuclear stellar disk and obtained the age of the Galactic bar to be
greater than  8 Gyr.

Many  galaxies with a bar include outer elliptical resonance rings
forming  near the OLR of the bar. There are two types of the outer
rings:  rings $R_1$ elongated perpendicular to the bar and rings
$R_2$ elongated parallel to the bar. The outer ring $R_1$ lies a bit
closer to the galactic center than the ring $R_2$ \citep{schwarz1981,
buta1991, byrd1994, buta1995, buta1996, rautiainen1999,
rautiainen2000}. The outer rings are supported by stable periodic
orbits which are followed by numerous quasi-periodic orbits
\citep{contopoulos1980, contopoulos1989}.

Data from the {\it Gaia} satellite give a unique opportunity to study
the motions of stars in a wide solar neighborhood \citep{prusti2016,
katz2018, brown2021, lindegren2021, vallenari2023}.

Many stars in the Galaxy move in orbits that lie both inside and
outside the resonances. \citet{weinberg1994} showed that there are
two types of orbits near the  ILR and OLR: orbits that change their
orientation relative to the bar in a certain range of angles and
orbits precessing without any angular restrictions.
\citet{struck2015a, struck2015b} showed that the eccentricity of
orbits slightly shifts the resonance positions. In the last decade,
study of stellar movements near the resonances has attracted special
attention \citep[][and other papers]{monari2017, trick2021,
chiba2021}

Studies of the kinematics and distribution of young stars  provided
evidence of the presence of a double outer ring $R_1R_2$ in the
Galactic disk \citep{melnikrautiainen2009, melnik2011,
rautiainen2010, melnik2015, melnik2016}. A comparison between model
velocities and velocities of OB associations in the Perseus,
Sagittarius and Local system star-gas complexes showed that the
angular velocity and position angle of the bar must be equal to
$\Omega_b=50\pm2$ km s$^{-1}$ kpc$^{-1}$ and
$\theta_b=40$--52$^\circ$, respectively \citep{melnik2019}.

Using {\it Gaia} EDR3 data, we built  the distribution of the
velocities of old disk stars along the Galactocentric distance $R$
and found that the best agreement between the model and observed
velocities corresponds to the angular velocity of $\Omega_b=55\pm3$
km s$^{-1}$ kpc$^{-1}$, position angle of $\theta_b=45\pm15^\circ$
and the age of the Galactic bar equal to $1.8\pm0.5$ Gyr
\citep{melnik2021}. We found periodic enhancement of either trailing
or leading segments of the resonance elliptical rings. In the region
of the inner ring, the period of morphological changes is
$P=0.57\pm0.02$ Gyr, which is very close to the revolution period
along the long-term orbits around the equilibrium points $L_4$ and
$L_5$.  In the region of the outer rings, the period of morphological
changes is $P=2.0\pm0.1$ Gyr which is probably related to the orbits
trapped into librations near the OLR \citep{melnik2023}.

The solar Galactocentric distance is  adopted to be   $R_0=7.5$ kpc
\citep[][]{glushkova1998, nikiforov2004, eisenhauer2005, bica2006,
nishiyama2006, feast2008, groenewegen2008, reid2009b, dambis2013,
francis2014, boehle2016, branham2017, iwanek2023}. On the whole, the
choice of the value of $R_0$ within the limits 7--9 kpc has
practically no effect on our results.

In this paper we study the formation of the humps (elevations) on the
profiles of the $V_R$-velocity distribution  along the Galactocentric
distances $R$. The period of the hump formation  is $P\approx 2$ Gyr,
and the height of the humps decreases with time. We  show that the
formation of the humps is connected with orbits trapped into
libration near the OLR which change their orientation with a period
of $P\approx 2$ Gyr.

The paper is organized as follows. Section 2 describes  the dynamic
model. The formation  of the humps on the profiles of the
$V_R$-velocity distribution is discussed in Section 3. The sample of
stars creating the humps is considered in Section 4. Section 5 gives
the examples of orbits supporting the humps.  The order of symmetry
and orientation of orbits as well as the fraction of librating orbits
near the OLR are discussed in Section 6. Section 7 studies the
distribution of stars in the plane ($E_J$, $P$), where $E_J$ is the
Jacobi integral and $P$ is the period of long-term oscillations in
the angular momentum and energy. The contribution of hump-creating
stars into the oscillations of the velocities $V_R$ and $V_T$ is
studied in Section 8. Section 9 presents the distribution of stars of
the model disk over the period $P$. The age of the Galactic bar is
discussed in Section 10. The main conclusions are given in Section
11.

\section{2. Model}

We use a 2D model of the Galaxy with an analytical Ferrers bar
\citep{freeman1972, athanassoula1983, pfenniger1984, sellwood1993,
binney2008} which reproduces well the distributions of the radial,
$V_R$, and azimuthal, $V_T$, velocities along the Galactocentric
distance $R$ derived from {\it Gaia} EDR3 and {\it Gaia} DR3 data.
The semi-major and semi-minor axes of the bar are $a=3.5$ and
$b=1.35$ kpc. The angular velocity of the bar is $\Omega_b=55$ km
s$^{-1}$ kpc$^{-1}$ which corresponds to the positions of the CR and
OLR at the distances $R_{RC}=4.04$ and $R_{OLR}=7.00$ kpc. The
ultraharmonic resonance $-4/1$ is located at the distance of
$R_{-4/1}=5.52$ kpc. The mass of the bar is $1.2 \times 10^{10}$
M$_\odot$.  At the initial moment the model disk is axisymmetric and
the non-axisymmetric components of the bar grow slowly.  The bar
gains its full strength during  4 bar rotation periods which
corresponds to the time $T_g= 0.45$ Gyr.  This duration of the bar
growth is consistent with estimates obtained for disk-dominated
galaxies in N-body simulations \citep{rautiainen2000,rautiainen2010}.
During the growth, the mass of the bar is conserved. The bar strength
corresponding to its full power is $Q_b=0.3142$ which matches strong
bars \citep{block2001, buta2004, diaz-garcia2016}.

The position angle of the Sun relative to the direction of the bar
major axis  is adopted to be $\theta_\odot=-45^\circ$.  Since our
model has the order of symmetry $m=2$, then both values of the
position angle $\theta_\odot=-45$ and $135^\circ$ are equivalent.

The model includes an exponential disk with a characteristic scale of
$R_d=2.5$ kpc and a mass of $M_d=3.25 \times 10^{10}\,$M$_\odot$. The
total mass of the bar and disk is $4.45 \times 10^{10}\,$M$_\odot$,
which is consistent with estimates of the mass of the Galactic disk
\citep{shen2010, fujii2019}.

The model also includes a classical bulge with a mass of
$M_{bg}=5\times 10^{9}\,$M$_\odot$ \citep{nataf2017, fujii2019} and
the isothermal halo \citep{binney2008}.

The total rotation curve of the model Galaxy is flat on the
periphery. The angular velocity of the disk rotation at the distance
of the Sun is $\Omega_0=30.0$ km s$^{-1}$ kpc$^{-1}$ which
corresponds to the value derived from the analysis of the kinematics
of OB associations \citep{melnik2020}. A more detailed description of
the model is given in \citet{melnik2021}.

The model disk  includes $2\times 10^6$ massless particles. The time
of simulation is 6 Gyr. We increased the simulation time from 3  to 6
Gyr to show that the formation of the humps in the radial velocity
profile is a periodic process.

\section{3. Humps  on  the  profiles  of the   $V_R$-velocity
distributions}


\begin{figure*}
\includegraphics[width=0.8\textwidth]{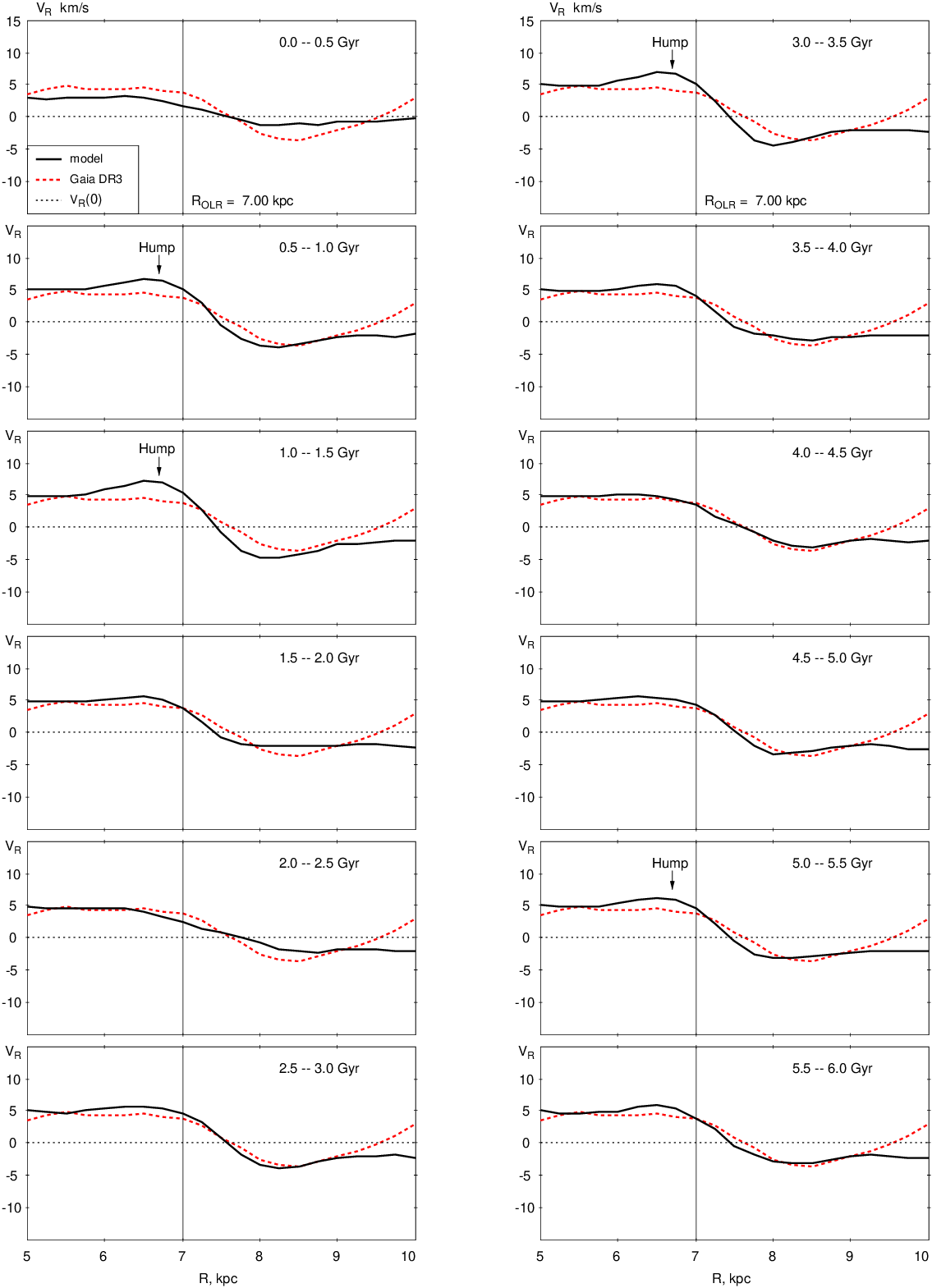}
\vspace{-0mm}     \centering \caption{Distributions of the model
radial velocities, $V_R$, along the Galactocentric distance, $R$,
averaged over the time intervals of 0.5 Gyr (black line). The
distribution of the observed velocity $V_R$ derived from the {\it
Gaia} DR3 data (red dashed line). For building the model and observed
profiles, we use stars located in a narrow sector of the azimuthal
angles, $|\theta-\theta_\odot|<15^\circ$. The median velocities $V_R$
are calculated  in  $\Delta R=250$-pc wide bins. Random errors in the
determination of the median velocities $V_R$ are smaller than the
line thickness. On the whole, the profiles show a plateau with
$V_R\approx 5$ km s$^{-1}$ in the distance range of 5--7 kpc followed
by a smooth fall to the value of $V_R \approx -3$ km s$^{-1}$ at the
distance of $R=8.5$ kpc and then a growth or a plateau. The model
velocity profiles form the humps (marked by arrows) at the distances
6--7 kpc at the time periods 0.5--1.0, 1.0--1.5, 3.0--3.5 and
5.0--5.5 Gyr.} \label{fig:_vr_profiles}
\end{figure*}

Figure \ref{fig:_vr_profiles} shows the  distributions of the model
radial velocities $V_R$ along the Galactocentric distance $R$
averaged over the time intervals of 0.5 Gyr (50 time instants
separated by 10 Myr). Also shown is the  distribution of observed
velocities derived from the {\it Gaia} DR3 data.  For building the
observed profiles, we used  $\sim 9.62\times10^6$ {\it Gaia} DR3
stars lying near the Galactic plane, $|z|<200$ pc, and in a narrow
sector of the azimuthal angles, $|\theta|<15^\circ$, with known
radial velocities and proper motions, the parallax-to-parallax error
ratio $\varpi/\varepsilon_\varpi>5$ and the error
$\textrm{RUWE}<1.4$. To construct the model profiles of the velocity
distribution, we used stars of the model disk lying in the sector
$|\theta-\theta_\odot|<15^\circ$ at different time instants. The
median velocities $V_R$ are calculated in $\Delta R=250$-pc wide
bins. The method of the determination of the median velocities and
their errors is described in \citet[][section 2]{melnik2021}. The
random error in the determination of the observed median velocities
$V_R$ calculated in bins in the distance range of 5--10 kpc is, on
average, 0.1 km s$^{-1}$. The similar error calculated for
instantaneous model velocities $V_R$ is 0.6 km s$^{-1}$ but it is
reduced to 0.1 km s$^{-1}$ after averaging over  time. On the whole,
the profiles show a plateau with $V_R\approx 5$ km s$^{-1}$ in the
distance range of 5--7 kpc, followed by a smooth fall to the value of
$V_R \approx -3$ km s$^{-1}$ at the distance of $R=8.5$ kpc and then
a growth or a plateau. We can clearly see that the model velocity
$V_R$ increases at the distances 6--7 kpc and forms the hump
(elevation) during the time periods 0.5--1.0, 1.0--1.5, 3.0--3.5 and
5.0--5.5 Gyr. The first appearance of the hump lasts long, from  0.6
to 1.8 Gyr, but next humps appear through $\sim 2$ Gyr and live $\sim
1$ Gyr. Besides, the height of the humps decreases with time.

We do not present  the distributions of the azimuthal velocities
$V_T$ here, although they also form humps, which are discussed below.

\begin{figure*}
\includegraphics[width=0.7\textwidth]{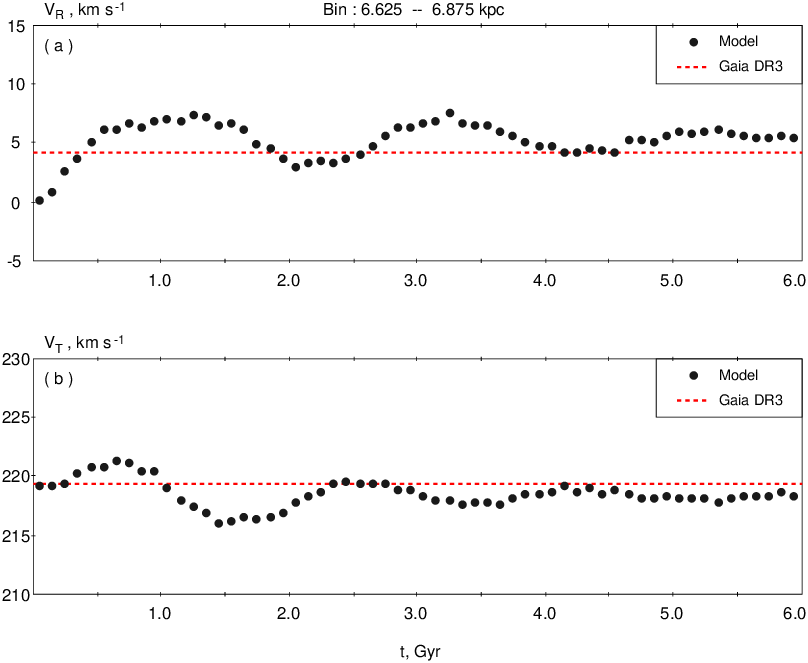}
\vspace{-0mm}     \centering \caption{Median radial,  $V_R$ (a), and
azimuthal, $V_T$ (b), velocities of stars of the model disk located
in the sector of the azimuthal angles
$|\theta-\theta_\odot|<15^\circ$ and in one distance bin
($6.625<R<6.875$ kpc)  at different times. The observed velocity
derived from the {\it Gaia} DR3 data (the red dashed line) in the
segment considered is also shown.} \label{fig:_vr_vt_var}
\end{figure*}

Figure \ref{fig:_vr_vt_var} shows the median radial, $V_R$, and
azimuthal, $V_T$, velocities of model stars in  the sector of the
azimuthal angles $|\theta-\theta_\odot|<15^\circ$ and in one distance
bin ($6.625<R<6.875$ kpc)  at different times. The middle of the bin
considered is located at the distance $R=6.75$ kpc. It is this bin
where the amplitude of the velocity variations achieves maximum
value. We consider all stars of the model disk that fall onto the
outlined segment  at different time instants. The median velocities
$V_R$ and $V_T$ are calculated for 601 time instants separated from
each other by 10 Myr and then are averaged over the periods of 100
Myr.

The variations of the velocities $V_R$  are approximated by harmonic
oscillations with the period $P$, amplitude $A$, initial phase
$\varphi$ and the average velocity $\overline{V_R}$. The parameters
of oscillations of the velocity $V_R$   are derived from the solution
of the system of equations:

\begin{equation}
V_{R,\,n}= \overline{V_R}+A\sin(2\pi t_n/P+\varphi),
 \label{vr_eq1}
\end{equation}

\noindent where $V_{R,\,n}$ are the velocity $V_R$ at the time
instants $t_n$, $n=0$, .., 60.  Eq.~(\ref{vr_eq1}) can be rewritten
in the following way:

\begin{equation}
V_{R,\,n}= \overline{V_R}+C_1 \cos(2\pi t_n/P)+C_2 \sin(2\pi t_n/P).
 \label{vr_eq2}
\end{equation}

\noindent  We use the standard least square method to solve the
system of 61 equations, which are linear in the parameters $C_1$,
$C_2$ and $\overline{V_R}$ for each value of the nonlinear parameter
$P$, and then determine the value of $P$ that minimizes the sum of
squared normalized  deviations $\chi^2$. The amplitude $A$ and phase
$\varphi$ are determined by the following expressions:

\begin{equation}
A= \sqrt{C_1^2+C_2^2}
 \label{A_eq}
\end{equation}

\noindent and

\begin{equation}
\varphi= \arctan \frac{C_1}{C_2}.
 \label{phi_eq}
\end{equation}

\noindent The uncertainty in  $A$  can be estimated with the use of
the uncertainties, $\varepsilon_1$ and $\varepsilon_2$, in the
parameters $C_1$ and $C_2$, respectively:

\begin{equation}
\varepsilon_A= \frac{\sqrt{C_1^2 \varepsilon_1^2+C_2^2
\varepsilon_2^2}}{A}.
 \label{varepsilon_A}
\end{equation}

%

\noindent  The parameters of the $V_T$-velocity oscillations are
calculated in a similar way.

Figure \ref{fig:_vr_vt_var}(a) shows the  median velocity $V_R$ of
stars lying in the outlined segment of the model disk at different
time moments. The average amplitude and  initial phase of the
$V_R$-velocity oscillations calculated over the 6 Gyr period is
$A=1.76\pm0.15$ km s$^{-1}$  and $\varphi=257\pm5^\circ$. The average
velocity $V_R$ at the 6 Gyr interval is $\overline{V_R}=5.2\pm0.1$ km
s$^{-1}$ which slightly exceeds the velocity derived from the {\it
Gaia} DR3 data, $V_R=4.15\pm0.04$ km s$^{-1}$. Maximum height of the
hump is $1.90\pm0.12$ km s$^{-1}$. The humps appear for the first,
second and third time during the time periods $t=0.6$--1.8 Gyr,
3.0--3.8 Gyr and 5.0--5.8 Gyr from the start of simulation,
respectively. The height of the humps decreases with time. The period
of the $V_R$-velocity variations  is $P=2.1\pm0.1$ Gyr.

Figure \ref{fig:_vr_vt_var}(b) shows variations of the  median
velocity $V_T$ of stars in the indicated segment of the model disk.
The average amplitude and  initial phase of the $V_T$-velocity
oscillations are $A=1.24\pm0.14$ km s$^{-1}$  and
$\varphi=329\pm6^\circ$. The average velocity  is
$\overline{V_T}=218.5\pm 0.1$ km s$^{-1}$ which is slightly less than
the value  derived from {\it Gaia} DR3 data, $V_T=219.30 \pm0.03$ km
s$^{-1}$. The height of the humps also decreases with time. The
period of  the $V_T$-velocity variations  is $P=1.9\pm0.1$ Gyr.

Thus, the velocities $V_R$ and $V_T$ calculated for the outlined
segment of the model disk, $|\theta-\theta_\odot|<15^\circ$ and
$R=6.75\pm0.125$ kpc, demonstrate  variations with the  period of
$P=2.0\pm0.1$ Gyr. The amplitudes of the velocity variations are 1--2
km s$^{-1}$ but their statistical significance  (the ratio of the
amplitude to its uncertainty) exceeds 8$\sigma$.

\section{4. Sample of stars creating the humps}

\subsection{4.1 Selection criterion}

We found stars that create the humps on the profiles of the
$V_R$-velocity distributions. Stars captured by the Lindblad
resonances often demonstrate the  periodic changes in the direction
of orbit elongation with respect to the bar major axis
\citep{weinberg1994}. We selected stars whose orbits are oriented in
such a way  that they create negative  velocities $V_R$ in the
region: $|\theta-\theta_\odot|<15^\circ$ and $R=6$--7 kpc, during the
certain time periods, and leave this region at other periods.

We calculated the angles $\theta_{01}$, $\theta_{02}$ and
$\theta_{03}$ which determine  orbit elongation relative to the bar
major axis during the time periods 0--1, 1--2 and 2--3 Gyr from the
start of modeling, respectively. Stars that change their orientation
as follows: $0<\theta_{01}<45^\circ$, $-45<\theta_{02}<0^\circ$ and
$0<\theta_{03}<45^\circ$ appear to contribute significantly in the
formation of the humps. Our sample includes 26308 stars, which is
only 9\% of all stars whose orbits lie both inside and outside the
OLR radius.

\begin{figure*}
\includegraphics[width=0.8\textwidth]{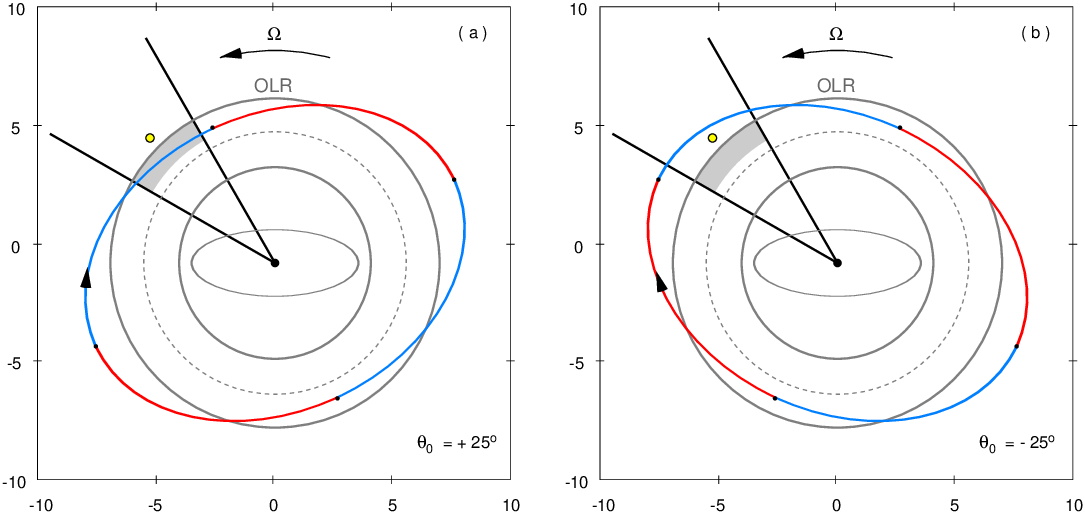}
\vspace{-0mm}     \centering \caption{Schematic representation of two
orbits oriented at the angles (a) $\theta_0=25^\circ$  and (b)
$\theta_0=-25^\circ$  to the major axis of the bar. The Galaxy
rotates counterclockwise, but in the reference system of the bar,
stars located beyond the CR rotate in the sense opposite that of the
Galactic rotation. It shows the segments of the orbits where stars
are approaching ($V_R<0$, blue lines) the Galactic center and moving
away ($V_R>0$, red lines) from it. The supposed position of the Sun
is shown by the yellow circle. Also shown is the sector of the
azimuthal angles $\theta=135\pm 15^\circ$ where the velocities are
calculated. The segment 6--7 kpc where the humps are forming is shown
in gray. Also shown are the bar (ellipse), the CR and OLR (solid gray
lines) and the resonance $-4/1$ (dashed gray line). We can clearly
see that a star moving along the orbit oriented at the angle
$\theta_0=25^\circ$ passes through the region where the humps are
forming  with   a negative radial velocity ($V_R<0$). Consequently,
such stars ``pull'' the median velocities $V_R$ towards more negative
values and produce pits. When the orbit is oriented at the angle
$\theta_0=-25^\circ$, the star passes   the Sun at the distance
$R>R_{OLR}$ and simply leaves the region  where the humps are
forming. } \label{fig:schema}
\end{figure*}

Figure \ref{fig:schema} shows two elliptical orbits oriented at the
angles (a) $\theta_0=25^\circ$  and (b) $\theta_0=-25^\circ$  to the
major axis of the bar. The change in the orientation of the orbits
causes the formation of the humps. The Galaxy rotates
counterclockwise, but in the reference system  rotating with the bar
angular velocity, stars located beyond the CR move clockwise. Orbital
segments where  stars are approaching ($V_R<0$) and moving away
($V_R>0$) from the Galactic center are shown by different colors. The
azimuthal angle of the Sun relative to the bar is supposed to be
$\theta_\odot=135^\circ$. Also shown is the sector of the azimuthal
angles $\theta=135\pm 15^\circ$  where the velocities are calculated.
The segment 6--7 kpc where the humps are forming is highlighted in
gray color. We can clearly see that  a star moving along the orbit
oriented at the angle $\theta_0=25^\circ$ (Fig.~\ref{fig:schema}a)
passes through the region where the humps are forming  with a
negative radial velocity ($V_R<0$). Consequently, such stars ``pull''
the median velocities $V_R$ towards more negative values and produce
pits. When the orbit is oriented at the angle $\theta_0=-25^\circ$
(Fig.~\ref{fig:schema}b), the star passes the Sun at the distance
$R>R_{OLR}$ and simply leaves the segment where the humps are
forming. It is the escape of stars with the negative velocities,
$V_R<0$, from the region considered that causes the formation of the
humps (see also Section~8).

If a star moves along the elliptical orbit oriented at the angle
$\theta_{0}=45^\circ$ to the  bar major axis,  it crosses the
Sun--Galactic center line with a zero radial velocity, $V_R=0$
(Fig.~\ref{fig:schema}). If the orbit is oriented at the angle
$0<\theta_{0}<45^\circ$ ($45<\theta_{0}<90^\circ$), then the star
passes the Sun with a negative (positive) radial velocity. The humps
appear to be created by stars moving in orbits oriented as follows:
$0<\theta_{01}<45^\circ$, $-45<\theta_{02}<0^\circ$ and
$0<\theta_{03}<45^\circ$ at the time periods 0--1, 1--2 and 2--3 Gyr,
respectively. In other words,  these orbits create pits at the time
periods 0--1 and 2--3 Gyr and their departure from the region
considered causes an increase in the median velocity $V_R$ and the
formation of the hump.

Note that the same requirement for the orientation of  orbits during
the time periods 0--1  Gyr ($0<\theta_{01}<45^\circ$) and 2--3 Gyr
($0<\theta_{03}<45^\circ$) favours selection of orbits with a period
of oscillations close, but not strictly equal, to 2 Gyr.

The requirement for the orbit orientation to be at the angle
$-45<\theta_{02}<0^\circ$ during the time period  1--2 Gyr  causes an
additional number of stars with negative radial velocities ($V_R<0$)
to appear just outside the OLR, $R>R_{OLR}$ (Fig.~\ref{fig:schema}b).
These stars ``pull'' the median radial velocity calculated in the
sector $\theta=135\pm 15^\circ$ towards more negative values. But the
median velocity $V_R$ demonstrates a smooth fall in the distance
range $R_{OLR}<R<R_{OLR}+1$ kpc (Fig.~\ref{fig:_vr_profiles}). So the
appearance of  additional stars with $V_R<0$ in this area makes this
fall sharper.

\subsection{4.2 Initial coordinates and velocities of stars creating the humps}

\begin{figure*}
\includegraphics[width=0.8\textwidth]{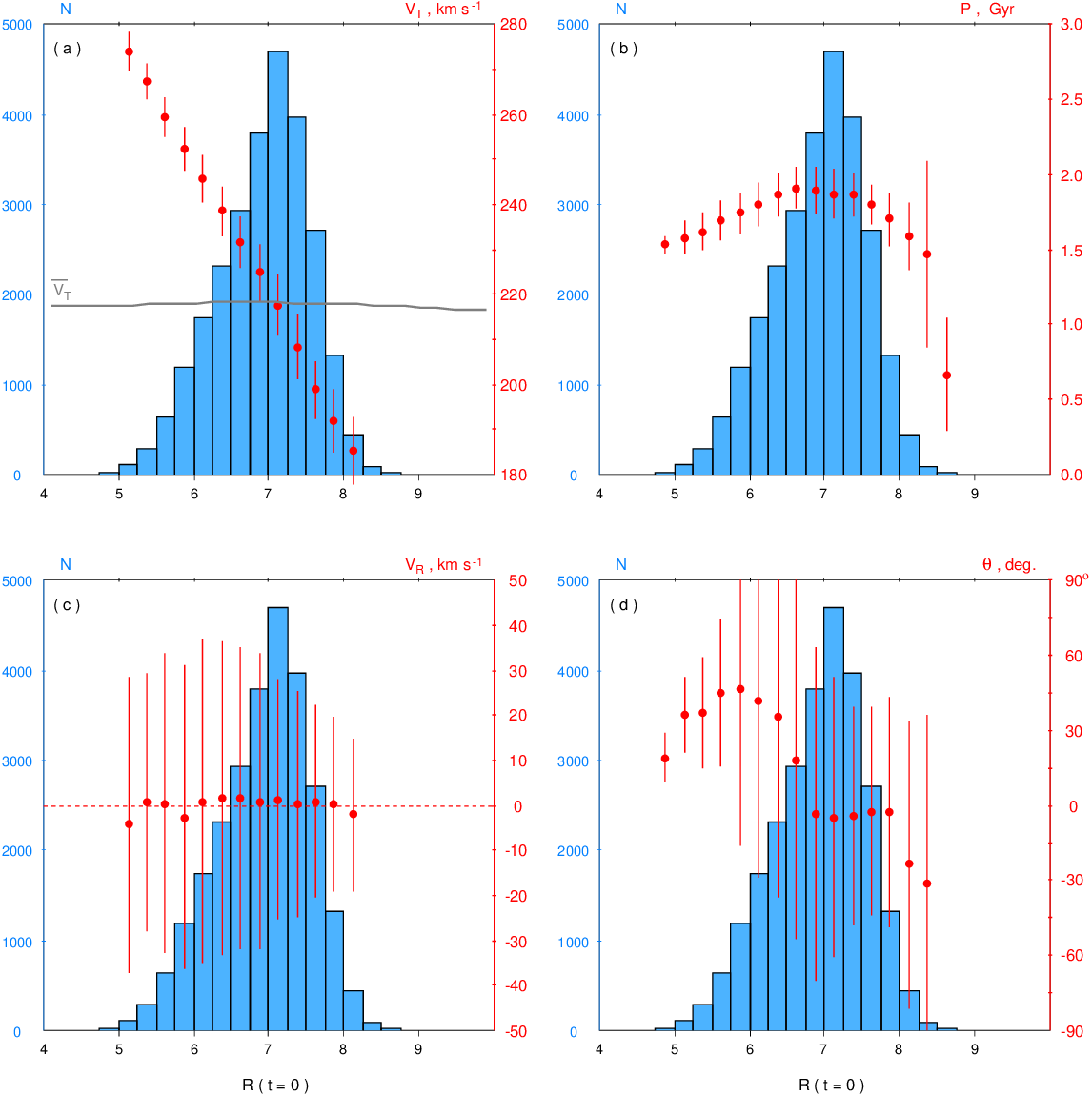}
\vspace{-0mm}     \centering \caption{Histograms of the distributions
of hump-creating stars over the distance $R$ at the initial time
moment (blue columns) with the distributions of other parameters (red
circles) superimposed upon them. In each bin over $R$, we calculated
(a) the median initial azimuthal velocity $V_T$, (b) the median
period $P$ of variations in the angular momentum and energy, (c) the
median initial radial velocity $V_R$, and (d) the median initial
azimuthal angle $\theta$. The red vertical lines show $+/-$
dispersion. (a) We can clearly see the decrease in $V_T$ with
increasing $R$. The gray line indicates the average azimuthal
velocity $\overline{V_T}$ of all stars located at a given distance
bin at $t=0$. The intersection of the gray line and the line drawn
through the red circles corresponds to the bin $R=7.00$--7.25 kpc,
where the majority of hump-creating stars are located at $t=0$. (b)
The median period $P$ calculated for the distance interval
$R=5.75$--8.00 kpc lies in the range $P=1.70$--1.91 Gyr. (c) The
median values of the initial radial velocity $V_R$ of hump-creating
stars is close to zero with the dispersion of $\sim 30$ km s$^{-1}$.
(d) The median azimuthal angle $\theta$ characterizing the stellar
position at $t=0$ changes jerkily from +45 to $-32^\circ$ in the
interval $R=4.0$--8.5 kpc, but it is close to zero, $\theta\approx
0$, near the OLR.} \label{fig:his_rg_2}
\end{figure*}

Figure \ref{fig:his_rg_2} shows the different dependencies obtained
for the initial coordinates and velocities of stars creating the
humps. It presents the  histograms of the star distribution over the
distance $R$ at the initial time moment ($t=0$) with the
distributions of other parameters  superimposed upon them. In each
bin over $R$, we calculated (a) the median  initial azimuthal
velocity $V_T$, (b) the median period $P$ of variations of the
angular momentum and energy, (c) the median initial radial velocity
$V_R$,  and (d) the median initial azimuthal angle $\theta$. The red
vertical lines show median values of dispersion $\sigma$ for the
corresponding quantities (half central interval containing 67\% of
objects).

Figure \ref{fig:his_rg_2}(a) shows the dependence of the initial
azimuthal velocity, $V_T$, from the distance $R$. We can clearly see
that $V_T$ decreases with increasing $R$. The gray line shows the
average initial azimuthal velocity $\overline{V_T}$ which is
calculated from the Jeans equation and is related to all stars
located at a given distance at $t=0$ \citep[][section
3.2]{melnik2021}. Due to asymmetric drift, the velocity
$\overline{V_T}$ is everywhere less than the velocity of the rotation
curve, $V_c$, and at the OLR radius, their difference amounts to
$\overline{V_T}- V_c= -7$ km s$^{-1}$. The intersection of the gray
line and the line drawn through the red circles corresponds to the
bin $R=7.00$--7.25 kpc, where the majority of stars creating the
humps lie at the initial moment.

In general, the decrease in $V_T$ with increasing $R$
(Fig.~\ref{fig:his_rg_2}a) is easy to understand. Stars creating the
humps must cross the OLR radius. So stars located  at the smaller
(greater) distance than the OLR radius at $t=0$ must ``go up'' (``go
down'') to the OLR radius. So they must have on average greater
(smaller) angular momentum, and therefore greater (smaller)
azimutulal velocity $V_T$, than other stars at the corresponding
distance.

Figure \ref{fig:his_rg_2}(b) shows changes in the period $P$ of
variations in the angular momentum  and energy  obtained for
hump-creating stars (see also Section~5). The median period $P$
calculated in bins at the distance interval $R=5.75$--8.00 kpc lies
in the range $P=1.70$--1.91 Gyr. The median period $P$ calculated for
all hump-creating stars  is 1.85 Gyr.

Figure \ref{fig:his_rg_2}(c) shows the distribution of the median
initial radial velocities, $V_R$, of  hump-creating stars. It is
clearly seen that the median velocities $V_R$ are close to zero and
lie in the range [$-3$, +2] km s$^{-1}$ at the distance interval
$R=5.75$--8.50 kpc. As for the dispersion of the radial velocities,
it is $\sim30$ km s$^{-1}$ and decreases with increasing distance
$R$, which is consistent with the general tendency for the velocity
dispersion to decrease on the periphery.

Figure \ref{fig:his_rg_2}(d) shows the median values of the initial
azimuthal angles $\theta$ calculated in each bin over the distance
$R$. We can clearly see that the angle $\theta$ changes jerkily from
+45 to $-32^\circ$ at the distance interval $R=4.0$--8.5 kpc, but
near the OLR ($R_{OLR}=7.00$ kpc), the median angle $\theta$ is close
to zero.

\section{5. Examples of orbits supporting the humps}

\label{examples}

\begin{figure*}
\includegraphics[width=0.8\textwidth]{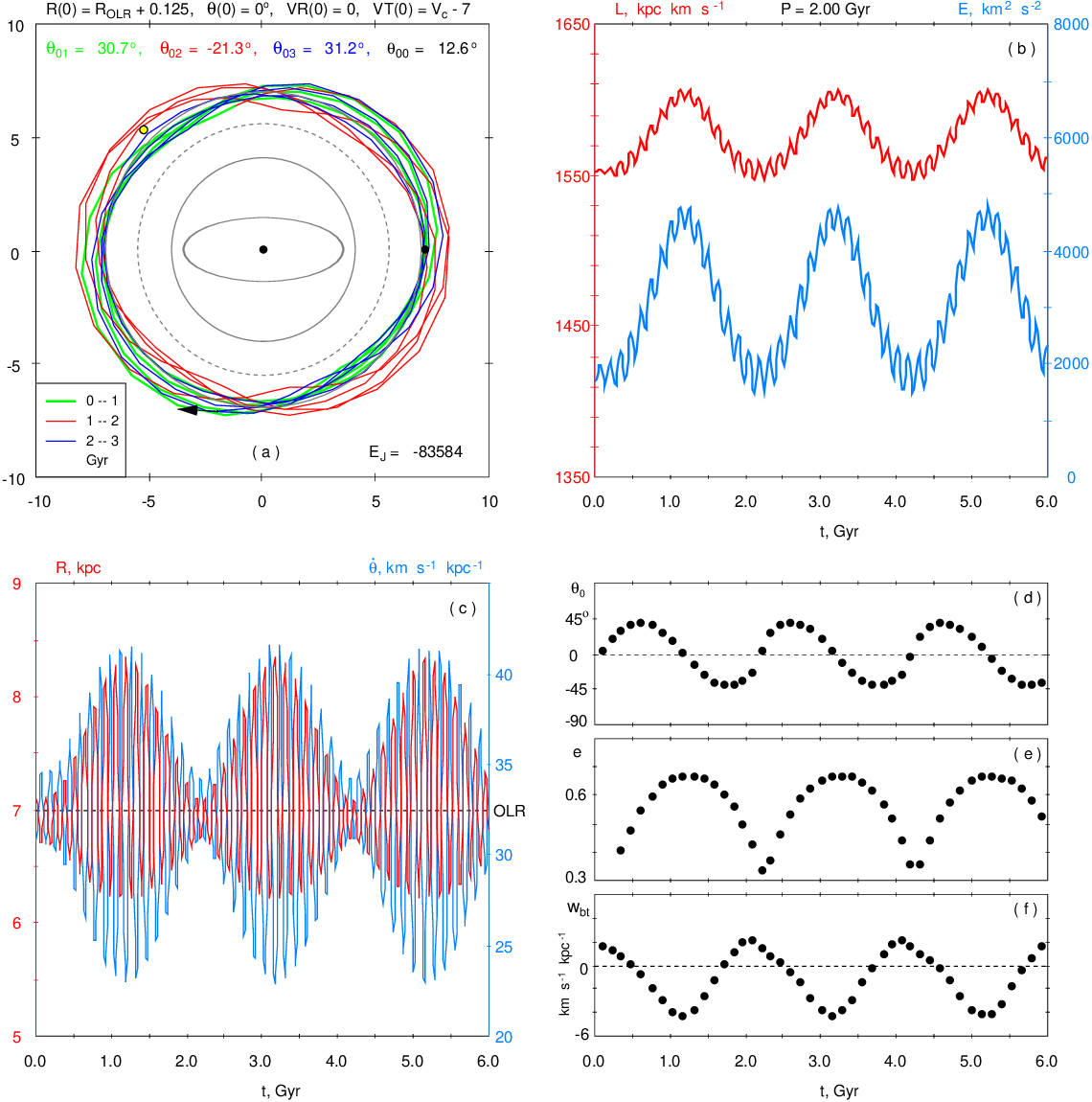}
\vspace{-0mm}     \centering \caption{Example of a typical orbit
supporting the humps. (a) The orbit is shown in the reference frame
of the rotating bar, in which the star considered rotates clockwise.
The supposed position of the Sun is shown by the yellow circle. At
the initial moment, the star is located at a distance $R=7.125$ kpc
in the direction of the major axis of the bar, $\theta=0^\circ$
(black circle), and has the radial and azimuthal velocities of
$V_R=0$ and $V_T=V_c-7$ km s$^{-1}$. The orbit of the star at the
time periods 0--1, 1--2 and 2--3 Gyr are shown in green, red and
blue, respectively. The angles $\theta_{01}$, $\theta_{02}$,
$\theta_{03}$ and $\theta_{00}$ determining the orbital orientation
relative to the major axis of the bar for the time periods 0--1,
1--2, 2--3 Gyr and over the entire period 0--3 Gyr are represented in
green, red, blue and black, respectively. We can clearly see that the
orbit of the star is tilted to the right (in the direction opposite
that of the Galactic rotation) during the time periods 0--1 and 2--3
and to the left during the period 1--2 Gyr. Also shown are the bar
position (ellipse), CR and OLR (solid gray lines), resonance $-4/1$
(dashed gray line) and   $E_J$. Distances, velocities and $E_J$ are
presented in units of kpc, km s$^{-1}$ and km$^2$  s$^{-2}$,
respectively. (b) Variations in the angular momentum $L$ (red line)
and total energy $E$ (blue line) of the star during the time period
0--6 Gyr.  We can clearly see the presence of short- and long-term
oscillations in $L$ and $E$. The period of long-term variations in
$L$ and $E$ is $P=2.0$ Gyr. The scales of changes in $L$ and $E$ are
shown on the left and right vertical axes, respectively. (c)
Variations  in the distance $R$ (red line) and instantaneous angular
velocity $\dot{\theta}$ (blue line). The OLR radius is shown by the
dashed line. It is clearly seen that variations in $R$ and
$\dot{\theta}$ have the form of beats. The scales of changes in $R$
and $\dot{\theta}$ are shown on the left and right vertical axes,
respectively. (d) Variations of the angle $\theta_0$, which
determines the direction of orbit elongation  at the period of one
radial oscillation, as a function of time. We can clearly see that
the angle $\theta_0$ changes slowly from +45 to $-45^\circ$ and then
quickly  back.  (e) Variations  in the orbital eccentricity, $e$,
calculated at the interval of one radial oscillation. (f) Variations
in the beat frequency, $w_{bt}$, which takes both positive and
negative values. } \label{fig:orb_1}
\end{figure*}

Figure \ref{fig:orb_1}(a) shows a typical  orbit supporting the
humps. The orbit is shown in the reference frame of the rotating bar
in which the star considered rotates in the sense opposite that of
Galactic rotation. The supposed position of the Sun is shown by the
yellow circle. At the initial moment the star is located at the
distance of $R=R_{OLR}+0.125$ kpc in the direction of the major axis
of the bar (black circle) and has radial and azimuthal velocities of
$V_R=0$ and $V_T=V_c-7$ km s$^{-1}$, which are the most probable
values of the velocities $V_R$ and $V_T$ at the given distance at
$t=0$. The angles $\theta_{01}=30.7\pm2.1^\circ$,
$\theta_{02}=-21.3\pm2.3^\circ$ and $\theta_{03}=31.2\pm2.7^\circ$
determine the orientation of the orbit relative to the major axis of
the bar during the time periods 0--1, 1--2 and 2--3 Gyr,
respectively, and satisfy the requirement: $0<\theta_{01}<45^\circ$,
$-45<\theta_{02}<0^\circ$ and $0<\theta_{03}<45^\circ$, which is
required for a star to be included in the sample of hump-creating
stars. We can clearly see that the orbit of the star is tilted to the
right (clockwise) during the time periods 0--1 and 2--3 Gyr and to
the left during the period 1--2 Gyr.

The angle $\theta_{00}=12.6\pm2.9^\circ$ determines the orientation
of the orbit during the entire time period 0--3 Gyr
(Fig.~\ref{fig:orb_1}a). Since $\theta_{00}$ is small, i.~e.
$\theta_{00}<15^\circ$,  we can conclude that the orbit considered
generally supports the outer ring $R_2$ stretched parallel to the
bar.

In barred galaxies, neither the angular momentum of a star, $L$, nor
its total energy, $E$, are conserved, but in the case of a stationary
bar, their linear combination  is  conserved in the form of the
Jacobi integral $E_J$ \citep[for example,][]{binney2008}:

\begin{equation}
E_J= E-\Omega_b L.
 \label{jacobi}
\end{equation}

\noindent  In our model, the Jacobi integral keeps its value after
the bar reaches its full power, $t>T_g$, where $T_g=0.45$ Gyr. The
star considered has $E_J=-83584$ km$^2$ s$^{-2}$, which is saved  up
to the last digit at $t>T_g$.

Figure \ref{fig:orb_1}(b) shows variations in the specific angular
momentum, $L$, and specific total energy, $E$, of the star considered
during the time period 0--6 Gyr. We can clearly see the short- and
long-term oscillations in $L$ and $E$ with the periods of $P=130\pm
10$ Myr and $P=2000 \pm 20$ Myr, respectively. The short-term
oscillations arise for all stars and occur twice during a period of
revolution of a star relative to the bar. The long-term oscillations
have a larger amplitude and arise for stars near the resonances.

Figure \ref{fig:orb_1}(c) shows variations in the distance $R$ and
instantaneous angular velocity $\dot{\theta}$. It is  clearly seen
that the variations in $R$ and $\dot{\theta}$ have the form of beats,
which are characterized by periodic  changes in  amplitudes of
oscillations. The star considered crosses the OLR radius at each
radial oscillation but the average values of $R$ and $\dot{\theta}$
also demonstrate slow oscillations. Variations in $R$ and
$\dot{\theta}$ occur in antiphase.

We divided  oscillations of the star into the time intervals from one
intersection  of the OLR radius with the negative radial velocity,
$V_R<0$, to another, and calculated  the average values of
$\overline{R}$ and $\overline{\dot{\theta}}$.   An example of such
oscillations of $\overline{R}$ and $\overline{\dot{\theta}}$ can be
found in \citet[][Fig. 11e]{melnik2023}.

Figure \ref{fig:orb_1}(d) shows variations in the angle $\theta_0$
which determines the direction of the orbit elongation  at the time
interval of one radial oscillation. The angle $\theta_0$ is measured
from the direction of the bar major axis. It is clearly seen that the
angle $\theta_0$  slowly changes from +45 to $-45^\circ$ and then
quickly back.

Figure \ref{fig:orb_1}(e) shows  variations in the orbital
eccentricity, $e$, calculated at the interval of one radial
oscillation. The eccentricity takes values in the range
$e=0.34$--0.67. Note that minimum eccentricity corresponds to minimum
average distance $\overline{R}$.

Figure \ref{fig:orb_1}(f) shows variations in the angular frequency
of beats, $w_{bt}$, calculated at the intervals of one radial
oscillation. The beat frequency is determined from the relation:

\begin{equation}
w_{bt}= \kappa(\overline{R})+2(\overline{\dot{\theta}}-\Omega_b),
 \label{w_bt}
\end{equation}

\noindent where $\overline{R}$ and $\overline{\dot{\theta}}$ change
with time. The beat frequency, $w_{bt}$, is a particular case of the
frequency of the resonance \citep[e.g.][]{chiba2021}. For calculation
of the epicyclic frequency, $\kappa$, we took into account the
correction related to orbital eccentricity \citep{struck2015a,
struck2015b}.  Maximum correction to $\kappa$ is 0.59 km s$^{-1}$
kpc$^{-1}$ which is small compared to the $\kappa$ values,
$\kappa(\overline{R})=43.5$--46.4 km s$^{-1}$ kpc$^{-1}$. It is  seen
that $w_{bt}$ takes both positive and negative values which means the
change of the sense of the angle $\theta_0$ shift (see also
Section~6.2).

\begin{figure*}
\includegraphics[width=0.8\textwidth]{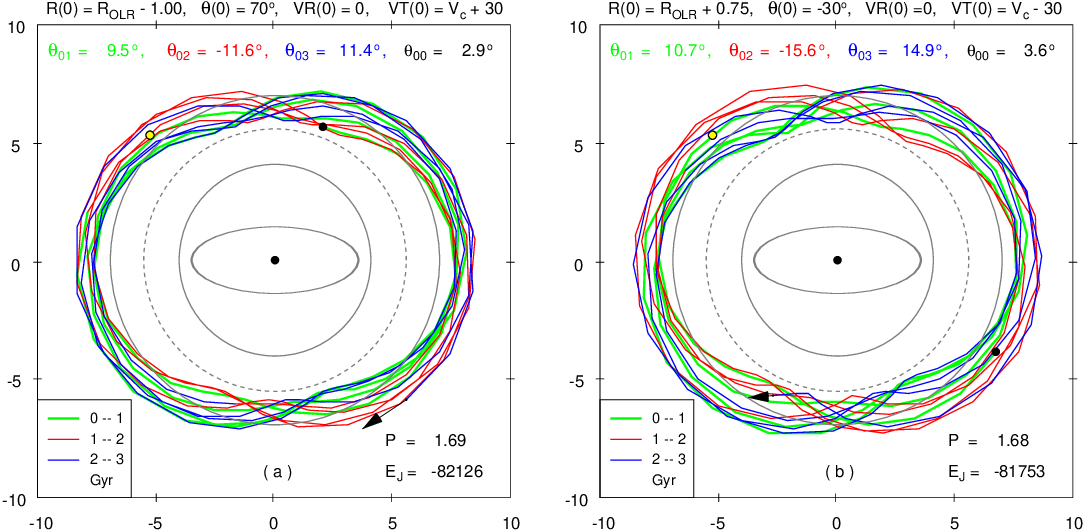}
\vspace{-0mm}     \centering \caption{Other examples of hump-creating
orbits. Orbits are shown in the reference frame of the rotating bar.
The supposed position of the Sun is shown by a yellow circle. (a) At
the initial moment, the star is located at the distance of
$R=R_{OLR}-1.00$ kpc and at the angle $\theta=70^\circ$ (black
circle) with the radial and azimuthal velocities of $V_R=0$ and
$V_T=V_c+30$ km s$^{-1}$. (b) The star has the following initial
coordinates and velocities: $R=R_{OLR}+0.75$ kpc, $\theta=-30^\circ$,
$V_R=0$ and $V_T=V_c-30$ km s$^{-1}$ (black circle). The orbits are
shown in green, red and dark blue during the time periods 0--1, 1--2
and 2--3 Gyr, respectively. The angles $\theta_{01}$, $\theta_{02}$,
$\theta_{03}$ and $\theta_{00}$ determine the orientation of the
orbits relative to the major axis bar during the time periods 0--1,
1--2, 2--3 Gyr and throughout the entire interval 0--3 Gyr,
respectively. Distances, velocities, $E_J$ and period $P$ are
presented in units of kpc, km s$^{-1}$, km$^2$  s$^{-2}$ and Gyr,
respectively. For more details, see caption to Fig.~\ref{fig:orb_1}.}
\label{fig:orb_2}
\end{figure*}

Figure \ref{fig:orb_2} shows two additional examples of hump-creating
orbits. Orientations of both orbits during the time periods 0--1,
1--2 and 2--3 Gyr satisfy the criterion $0<\theta_{01}<45^\circ$,
$-45<\theta_{02}<0^\circ$ and $0<\theta_{03}<45^\circ$. At the
initial moment, both stars are located at some distance away  from
the radius OLR: the first star (a) lies closer to the Galactic
center, $R=R_{OLR}-1.00$ kpc, while the second star (b) lies further
away from the center, $R=R_{OLR}+0.75$ kpc. To satisfy the above
criterion, the initial azimuthal velocities of these stars must
differ strongly from the velocity of the rotation curve, $V_c$. Note
that the first (second) star has the initial velocity  by 30 km
s$^{-1}$ higher (smaller) than the velocity of the rotation curve  at
the corresponding  distance which reflects the general trend
(Fig.~\ref{fig:his_rg_2}a).

Figure \ref{fig:orb_2} also shows  the periods of variations in the
angular momentum and energy of the stars considered which amount to
$P=1.69$ and 1.68 Gyr. Note that these values are a bit smaller than
the period of $P=2.00$ Gyr found for the star shown in Figure
\ref{fig:orb_1}. This difference also reflects the general trend: the
median period $P$ decreases as the initial distance of a stars moves
away from the OLR radius (Fig.~\ref{fig:his_rg_2}b).

Figure \ref{fig:orb_2} also shows the values of the Jacobi integral
of the stars considered: $E_J=-82126$ and $-81753$ km$^2$  s$^{-2}$
which have a bit larger values than $E_J=-83584$ km$^2$  s$^{-2}$ of
the star shown in Figure \ref{fig:orb_1} (see also Section~7).

Note that the choice of the initial values of the radial velocity
$V_R$ and azimuthal angle $\theta$  also affect the period $P$ and
average orientation of the orbit.

\section{6. Statistics of orbits near the OLR}

\subsection{6.1 Order of symmetry  and orientation}

We selected 289062 stars whose orbits lie  both inside and outside
the OLR radius and do not intersect the CR  and determined their
order of symmetry  $m$ and, if it is an elliptical orbit ($m=2$), the
direction of the orbit elongation.

To determine the order of symmetry of a stellar orbit, we used the
following system of equations:

\begin{equation}
R_n= R_{00}+\cos(m(\theta_n-\overline{\theta})),
 \label{R_n_1}
\end{equation}

\noindent where $R_n$ and $\theta_n$ are the Galactocentric distance
and the azimuthal angle of the star relative to the major axis of the
bar at time moments $t_n=10 n$ Myr, where $n=0$,.., 300. The values
$R_{00}$ and $\overline{\theta}$ determine the average Galactocentric
distance and average orientation  during the time period 0--3 Gyr.
This system of equations can be rewritten as follows:

\begin{equation}
R_n= R_{00}+C_1 \cos(m\theta_n)+C_2 \sin(m\theta_n),
 \label{R_n_2}
\end{equation}

\noindent where the angle $\overline{\theta}$ is derived from the
following expression:

\begin{equation}
\overline{\theta}= \frac{1}{m}\arctan \frac{C_2}{C_1}
 \label{R_n_2}
\end{equation}

\noindent and characterizes the average direction of the orbit
elongation relative to the major axis of the bar during the time
period $t=0$--3 Gyr. The angle $\overline{\theta}$ lies in the range
$-90 \le\overline{\theta}<90^\circ$ and is related to the angle
$\theta_{00}$ (Section~5) as follows:

\begin{equation}
\theta_{00}=\overline{\theta}+180^\circ \label{R_n_3}.
\end{equation}

We considered the following values of the order of symmetry: $m=2$,
3, 4 and 5, and calculated the corresponding values of the function
$\chi^2$($m$). The value of $\sigma_0$ was adopted to be $\sigma_0=1$
kpc for all $m$. If one  of the four $\chi^2$($m$), for example
$\chi^2$($m_1$), took a value less than other three values by some
critical value $\chi^2_c$:

\begin{equation}
\left \{
\begin{array}{l}
\chi^2(m_1)< \chi^2(m_2)- \chi^2_c \\
\chi^2(m_1)< \chi^2(m_3)- \chi^2_c \\
\chi^2(m_1)< \chi^2(m_4)- \chi^2_c, \\
\end{array}
\right.\\
\label{R_n_3}
\end{equation}

\noindent then the order of symmetry of the orbit was chosen to be
$m_1$. If this condition is not satisfied for any $m$ then we
believed that the shape of the orbit is close to circular, i.e.
$m=0$. The critical value $\chi^2_c$ was adopted to be $\chi^2_c=2$
which allows us to select orbits with a certain shape and exclude
combined orbits with elements  corresponding to different values of
$m$.

Table \ref{tab:statistics} lists the number of orbits with the order
of symmetry $m=0$, 2, 3, 4 and 5 and their fraction from the total
sample. For $m=2$, it also gives the number of orbits oriented in the
following sectors: $0\le\theta_{00}<15^\circ$,
$15\le\theta_{00}<75^\circ$, $75\le\theta_{00}<105^\circ$,
$105\le\theta_{00}<165^\circ$ and $165\le\theta_{00}<180^\circ$. For
each sample of stars, we calculated the median  period $P$ and its
dispersion, $\sigma_P$. Table~\ref{tab:statistics} consists of two
parts: Part I studies stars whose orbits lie both inside and outside
the OLR (289062 stars) and Part II considers the hump-creating stars
(26308 stars).

Table \ref{tab:statistics} (Part I) shows that the majority of
orbits, 33.4\%, lying both inside and outside the OLR  are oriented,
on average,  perpendicular to the major axis of the bar,
$75\le\theta_{00}<105^\circ$, i.~e are elongated along the minor axis
of the bar. However, the median period $P$ calculated for this stars
is $P=780$ Myr, therefore, they cannot create the humps with a period
of $P\approx 2000$ Myr. In addition, the dispersion of periods is
$\sigma_P=745$ Myr here, so this subset of stars cannot produce some
organized variations in orbital orientation for a long time.

Quite a lot of stars, 11.1\%, have orbits oriented at the  angles in
the range  $0\le\theta_{00}<15^\circ$ to the major axis of the bar.
The median period $P$ and its dispersion are $P=1730$ and
$\sigma_P=220$ Myr here. This sample of stars may well create the
humps with a period of $P\approx 2000$ Myr. The total fraction of
orbits oriented along the major axis of the bar, i.e. at the angle
$\theta_{00}$ in the range $0\le\theta_{00}<15^\circ \cup
165\le\theta_{00}\le 180^\circ$ is 16.6\%.

We paid special attention to orbits with the order of symmetry $m=3$,
as the resonance $-3/1$ ($R_{-3/1}=6.02$ kpc), like the OLR, lies
close to the radius where the humps are forming. Table
\ref{tab:statistics} (Part I) shows that the fraction of orbits with
$m=3$ is  7.7\% which is considerably smaller than the fraction of
orbits with  $m=2$ (54.8\%). The median  period $P$ for  orbits with
$m=3$ is $P=490$ Myr. Therefore, these stars cannot create humps with
a period of $P\approx 2000$ Myr.

The fraction of orbits with the order of symmetry $m=4$ is only 5.7\%
of all stars near the OLR (Table~\ref{tab:statistics}, Part I).
However, the median period is quite large, $P=1930$ Myr, here. On the
other hand, the dispersion of $P$ is quite large too, $\sigma_P=795$
Myr. Generally, these stars can create the humps with a period of
$P\approx 2000$ Myr but they must quickly dissolve with time.

For completeness, we considered the order of symmetry $m=5$. The
fraction of  these orbits is only 3.3\% and they do not play a
significant role in the hump-formation (Table~\ref{tab:statistics},
Part I).

A fairly large fraction of stars, 28.6\%, have orbits with the order
of symmetry $m=0$ (Table~\ref{tab:statistics}, Part I). This means
that none of the values of $m$ from the range $m=2$--5 describes
these orbits significantly better than others. The median  period and
dispersion are $P=720$ and $\sigma_P=590$ Myr here. So this sample of
stars cannot create oscillations with a period of $P\approx 2000$
Myr.

A different picture is observed for hump-creating stars (Table
\ref{tab:statistics}, Part II). Here 71.6\% of orbits are stretched
in the  range of angles $0\le\theta_{00}< 15^\circ$. These orbits
are, on average, elongated along the major axis bar and support the
ring $R_2$. The median period and dispersion are $P=1820$  and
$\sigma_P=155$ Myr here. It is clearly seen that this sample of stars
can support  the humps with a period of $P\approx 2000$ Myr for a
long time.

\begin{table*}
\caption{Orbit symmetry  and orientation} \centering
  \begin{tabular}{ccccc}
 \\[-7pt]\hline\\[-7pt]
 \multicolumn{5}{c}{Part I}\\
 \multicolumn{5}{c}{All orbits lying both in- and outside  OLR}\\
 $\theta_0$ & $N$ & $f$  & $P$ & $\sigma_P$ \\
  &  &   & Myr &  Myr \\
All & 289062 & 100.0\% & 910  & 715 \\
\\[-7pt]\hline\\[-7pt]
 \multicolumn{5}{c}{ m=2}\\
All & 158332 & 54.8\% & 1320  & 695 \\
$0 \le \theta_0 < 15^\circ$ & 32212 & 11.1\% & 1730  & 220 \\
$15 \le \theta_0 < 75^\circ$ & 10122 & 3.5\% & 2020  & 210 \\
$75 \le \theta_0 < 105^\circ$ & 96650 & 33.4\% & 780  & 745 \\
$105 \le \theta_0 < 165^\circ$ & 3316 & 1.1\% & 2030  & 560 \\
$165 \le \theta_0 < 180^\circ$ & 16032 & 5.5\% & 1720  & 270 \\
\\[-7pt]\hline\\[-7pt]
\multicolumn{5}{c}{ m=3}\\
All & 22226 & 7.7\% & 490  & 130 \\
\\[-7pt]\hline\\[-7pt]
\multicolumn{5}{c}{ m=4}\\
All & 16406 & 5.7\% & 1930  & 795 \\
\\[-7pt]\hline\\[-7pt]
\multicolumn{5}{c}{ m=5}\\
All & 9480 & 3.3\% & 650  & 80 \\
\\[-7pt]\hline\\[-7pt]
\multicolumn{5}{c}{ m=0}\\
All & 82618 & 28.6\% & 720  & 590 \\
\\[-7pt]\hline\\[-7pt]
\\[-15pt]\hline\\[-7pt]
 \multicolumn{5}{c}{Part II}\\
 \multicolumn{5}{c}{Orbits creating the humps}\\
 $\theta_0$ & $N$ & $f$  & $P$ & $\sigma_P$ \\
  &  &   & Myr &  Myr \\
All & 26308 & 100.0\% & 1850  & 170 \\
\\[-7pt]\hline\\[-7pt]
 \multicolumn{5}{c}{ m=2}\\
All & 23310 & 88.6\% & 1850  & 165 \\
$0 \le \theta_0 < 15^\circ$ & 18824 & 71.6\% & 1820  & 155 \\
$15 \le \theta_0 < 75^\circ$ & 4054 & 15.4\% & 1990  & 125 \\
$75 \le \theta_0 < 105^\circ$ & 0 & 0.0\% & --  & -- \\
$105 \le \theta_0 < 165^\circ$ & 0 & 0.0\% & --  & -- \\
$165 \le \theta_0 < 180^\circ$ & 432 & 1.6\% & 1460  & 40 \\
\\[-7pt]\hline\\[-7pt]
\multicolumn{5}{c}{ m=3}\\
All & 24 & 0.1\% & 650  & 5 \\
\\[-7pt]\hline\\[-7pt]
\multicolumn{5}{c}{ m=4}\\
All & 1764 & 6.7\% & 1810  & 150 \\
\\[-7pt]\hline\\[-7pt]
\multicolumn{5}{c}{ m=5}\\
All & 8 & 0.0\% & 1630  & 40 \\
\\[-7pt]\hline\\[-7pt]
\multicolumn{5}{c}{ m=0}\\
All & 1202 & 4.6\% & 1880  & 555 \\
\\[-7pt]\hline\\[-7pt]
\end{tabular}
\label{tab:statistics}
\end{table*}

\subsection{6.2 Fraction of librating orbits near the OLR}

\label{frac_lib}

We tried to separate orbits librating within certain angles from
orbits which have the direction of elongation shifting only in one
direction without any angular restriction. A necessary condition for
an orbit to be captured into libration is that it must lie both
inside and outside the radius of the resonance. For each of 289062
stars whose orbits lie both inside and outside the OLR, we calculated
the values of the beat frequency $w_{bt}$ (Eq.~\ref{w_bt}). Criterion
for orbital libration is following: maximum and minimum values of the
beat frequency $w_{bt}$ must be of different signs, $\max w_{bt}>0$
and $\min w_{bt}<0$, i.e. the beat frequency must change sign within
one oscillation. If  $w_{bt}$ does not change sign, then the
direction of orbit elongation shifts only in one direction: for
$w_{bt}>0$ -- in the sense of Galactic rotation and for $w_{bt}<0$ --
in the opposite sense. The fraction of orbits trapped into libration
near the OLR appears to be 28\% of orbits lying both inside and
outside the OLR.

\section{7. Distribution of stars in the plane ($E_J$,~$P$)}

\label{jacobi}

\begin{figure*}
\includegraphics[width=0.80\textwidth]{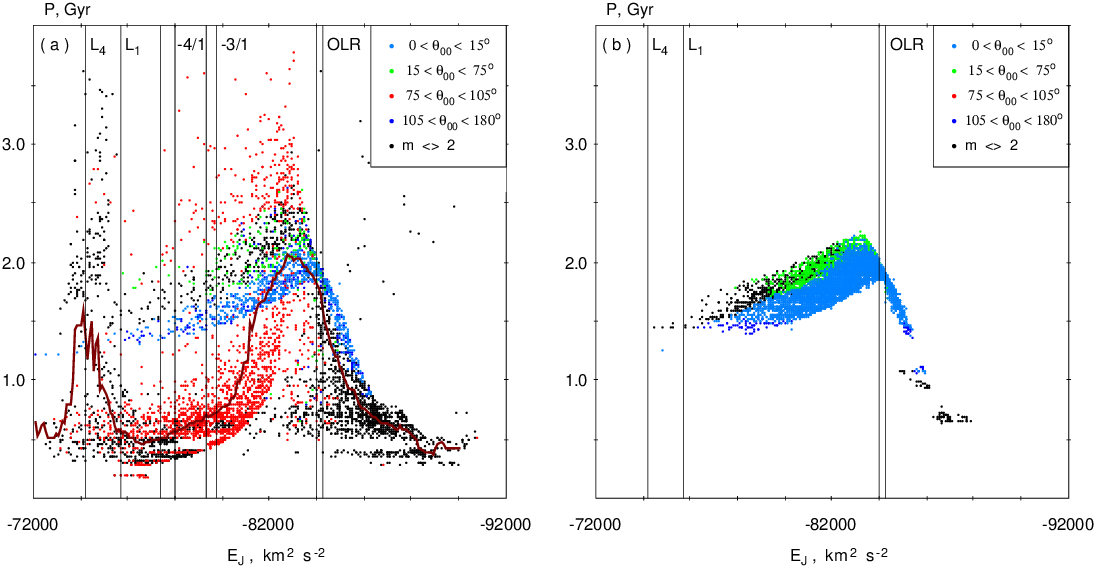}
\vspace{-0mm}     \centering \caption{Distributions of stars in the
plane ($E_J$, $P$), where $E_J$ is the Jacobi integral and $P$ is the
period of variations  in the angular momentum and energy. Orientation
of orbits with order of symmetry $m=2$ is shown in color:
$0\le\theta_{00}<15^\circ$ (blue), $15 \le\theta_{00}<75^\circ$
(green), $75 \le\theta_{00}<105^\circ$ (red), $105 \le
\theta_{00}<180^\circ$ (dark blue), where the angle $\theta_{00}$
determines the direction of elongation of the orbit relative to the
major axis of the bar during the time period 0--3 Gyr and lies in the
range of $0 \le\theta_{00}<180^\circ$. Orbits with $m \neq 2$ are
shown in black. The vertical lines show the values of $E_J$
calculated for imaginary stars located at points $L_1$ and $L_4$, and
at the radii of the resonances $-3/1$, $-4/1$ and OLR in the
directions of the major (more negative values) and minor axes of the
bar with the azimuthal velocities  equal to those of the rotation
curve at the corresponding radii. (a) All model stars whose orbits
lie both inside and outside the OLR (289062 stars). Only 2\% of stars
are shown. The solid red curve shows the median periods $P$
determined in each bin by $E_J$. It is clearly seen that the median
period $P$ increases near the OLR and then drops sharply. (b) Stars
creating the humps (26308 stars). 25\% of stars are shown. We can see
that the majority of stars have orbits oriented at the angle in the
range $0\le\theta_{00}<15^\circ$ (blue dots) which form a structure
resembling an angle: the median period $P$ grows almost linearly from
$P=1.6$ to 2.0 Gyr with decreasing $E_J$ from $-78000$ to $-83400$
km$^2$  s$^{-2}$ and then drops sharply.} \label{fig:jacobi}
\end{figure*}

Figure \ref{fig:jacobi} shows the distribution of stars in the plane
($E_J$, $P$), where $E_J$ is the Jacobi integral (Eq.~7) and $P$ is
the period of variations in the angular momentum and total energy.
Orientation of orbits with the order of symmetry $m=2$ is shown in
color: $0\le\theta_{00}<15^\circ$ (blue), $15\le\theta_{00}<75^\circ$
(green), $75\le\theta_{00}<105^\circ$ (red),
$105\le\theta_{00}<180^\circ$ (dark blue), where the angle
$\theta_{00}$ determines the direction of elongation of the orbit
relative to the major axis of the bar during the time period 0--3
Gyr. Orbits with the order of symmetry $m \neq 2$ are shown in black.

Figure \ref{fig:jacobi}(a) shows the distribution in the plane
($E_J$, $P$) built for  stars whose orbits lie both inside and
outside the OLR (289062 stars). To prevent the overload of the
drawing, we present only 2\% of stars chosen at random. The solid red
curve shows the median period  $P$ calculated in 100-km$^2$  s$^{-2}$
wide bins along $E_J$. The vertical lines show the values of $E_J$
calculated for imaginary stars located at the points $L_1$ and $L_4$
and at the radii of the resonances $-3/1$, $-4/1$ and OLR with the
velocities $V_R=0$ and $V_T=V_c$, where $V_c$ is the velocities of
the rotation curve at the corresponding radii. The double lines show
the values of $E_J$ calculated for stars located on the major (more
negative values) and minor axes of the bar. The values of $E_J$
calculated for the OLR radius are $E_J=-84266$ (major axis) and
$-83997$ km$^2$  s$^{-2}$ (minor axis). It is clearly seen that the
median period $P$ increases near the OLR and reaches the value of
$P=2.00$ Gyr and then quickly falls. Also, note a small increase in
the median period $P$ near the value $E_J=-74243$ km$^2$  s$^{-2}$
corresponding to the equilibrium point $L_4$.

Figure \ref{fig:jacobi}(b) shows the distribution of hump-creating
stars (26308 stars) in the plane ($E_J$, $P$). It is seen that the
majority of stars have orbits oriented at the angle
$0\le\theta_{00}<15^\circ$ (blue dots) and form a structure
resembling an angle: the median period $P$ increases practically
linearly from $P=1.60$ to 2.00 Gyr with decreasing $E_J$ from
$-78000$ to $-83400$ km$^2$  s$^{-2}$ and then quickly drops.

\section{8. Contribution of hump-creating stars to the
 oscillations  of the velocities $V_R$ and $V_T$}

 \label{contribution}

\begin{figure*}
\includegraphics[width=0.70\textwidth]{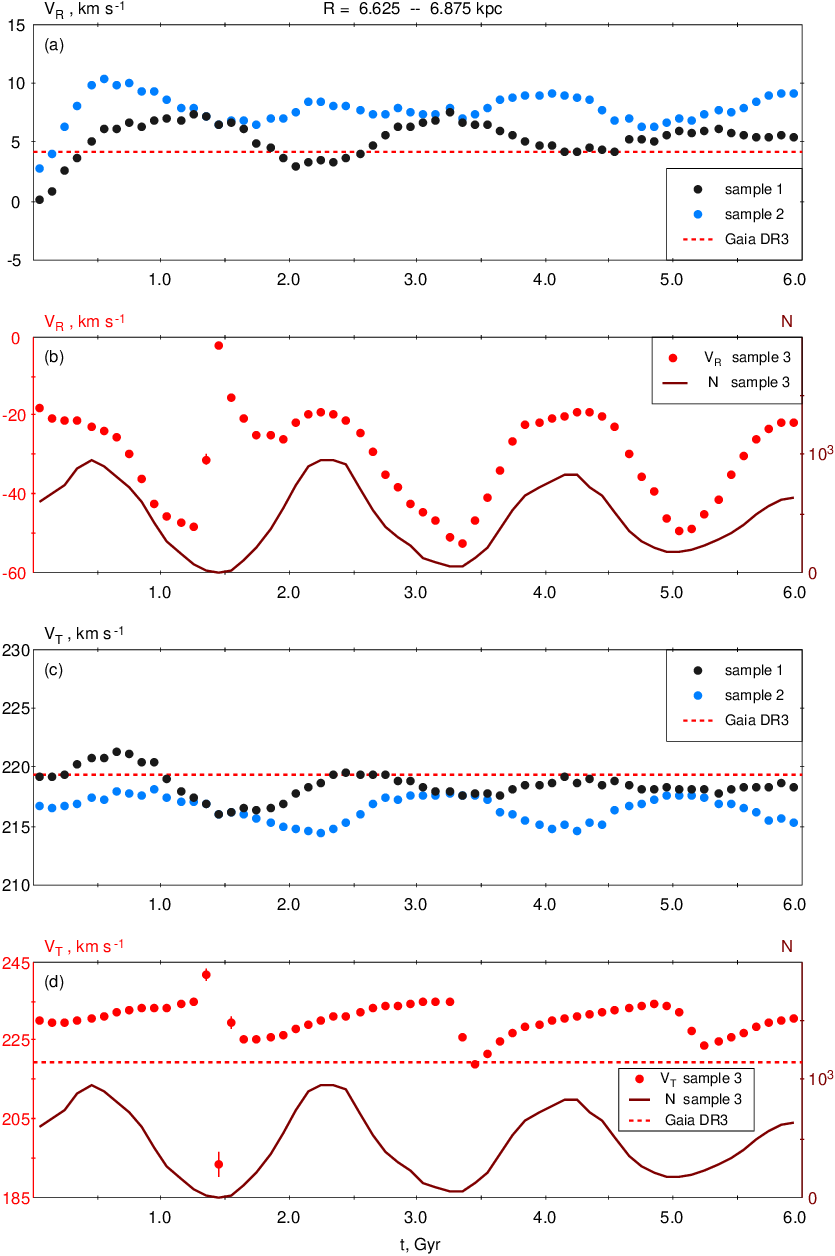}
\vspace{-0mm}     \centering \caption{Median (a, b) radial $V_R$ and
(c, d) azimuthal $V_T$ velocities of  model stars lying in the sector
$|\theta-\theta_\odot|<15^\circ$ and distance bin $R=6.75\pm0.125$
kpc  at different time instants. The velocities $V_R$ and $V_T$ are
calculated for three  samples of stars: (1) all stars of the model
disk (black circles), (2) all stars of the model disk without
hump-creating stars (blue circles), and (3) stars creating the humps
(red circles). The solid dark-red line shows variations in the number
of hump-creating stars, $N$, located in the outlined segment at
different times. The scale for variations in  $N$ is shown on the
right vertical axis. See also caption for Fig.~\ref{fig:_vr_vt_var}.
} \label{fig:vr_vt_var_long}
\end{figure*}

Figure \ref{fig:vr_vt_var_long} shows the median (a, b) radial $V_R$
and (c, d) azimuthal $V_T$ velocities of model stars lying in the
sector of the azimuthal angles $|\theta-\theta_\odot|<15^\circ$ and
distance bin $R=6.75\pm0.125$ kpc at different time instants. We
consider three samples of stars: (1) all stars of the model disk that
fall onto the outlined segment at the moments considered (black
circles), (2) all stars of the model disk without hump-creating stars
(blue circles), and (3) stars creating the humps (red circles). Also
shown are variations in the number of hump-creating stars, $N$,
located in the outlined segment of the disk at different times (solid
dark-red line).

Figure \ref{fig:vr_vt_var_long}(a) shows variations in the velocity
$V_R$ calculated for all stars of the model disk and for the sample
without hump-creating stars. We can clearly see that the exclusion of
hump-creating stars changed significantly  the phase of oscillations
of the velocity $V_R$: pits appear in places of the humps, except for
the first one, and humps appear in places of pits.


Let us consider a quantitative criterion characterizing the influence
of hump-creating stars on  the oscillations of the velocity $V_R$. A
good one is the ratio of the amplitude $A$ to  its uncertainty
$\varepsilon_A$. Note that the ratio $|A/\varepsilon_A|$ calculated
for  all stars located in the outlined segment is
$|A/\varepsilon_A|=11.7$ ($A=1.76\pm0.15$). However, the amplitude
$A$ computed without hump-creating stars but at the fixed values of
the period and phase, i.~e. at $P=2.1$ Gyr and $\varphi=257^\circ$,
which determine the positions of the humps and pits for the sample of
all stars, is $A=-0.21\pm0.19$ km s$^{-1}$. We can see that in this
case, the amplitude $A$ has a value close to its uncertainty and the
ratio is $|A/\varepsilon_A|=1.1$. Therefore, variations in the
velocity $V_R$ obtained for all stars located in the indicated
segment practically disappear after the exclusion of hump-creating
stars. However, we can see the appearance of other oscillations in
the velocity $V_R$ with humps and pits corresponding to other time
moments.

Figure \ref{fig:vr_vt_var_long}(b) shows variations in the velocity
$V_R$ and the number of stars $N$ obtained for the sample of
hump-creating stars (26308 stars). Let us pay attention to the range
of the velocity variations: from $-50$ to 0 km s$^{-1}$. Besides,
minimum velocities $V_R$ correspond to minimum values of $N$. In this
case, oscillations of the velocity  $V_R$  have the following
parameters: $P=2.0\pm0.1$ Gyr, $\overline{V_R}=-30.3\pm0.9$ km
s$^{-1}$, $A=13.1\pm1.7$ km s$^{-1}$  and $\varphi=75\pm5^\circ$.

A comparison between Fig.~\ref{fig:vr_vt_var_long}(a) and
Fig.~\ref{fig:vr_vt_var_long}(b) shows that the median velocities
$V_R$ of hump-creating stars (red circles) are always smaller than
the median velocities of all stars in the outlined segment (black
circles). The average velocities $V_R$ in the first and second case
equal $\overline{V_R}=-30.3$ and $\overline{V_R}=5.2$ km s$^{-1}$,
respectively. It means that hump-creating stars ``pull'' the median
velocities $V_R$ towards more negative values. When the number $N$
(solid dark-red line) of hump-creating stars increases, their
influence becomes more noticeable, and as a result, the median
velocities $V_R$ of all stars decrease. Thus, it is the periodic
variations in the number of hump-creating stars, $N$, that cause
oscillations of the velocity $V_R$ in the outlined segment.

Figure \ref{fig:vr_vt_var_long}(c) shows variations in the velocity
$V_T$ calculated for all stars located in the outlined segment (black
circles) and for the sample that does not include hump-creating stars
(blue circles). It is clearly seen that the exclusion of
hump-creating stars shifts, except for the first hump, the phase of
oscillations by $\sim 180^\circ$.


Figure \ref{fig:vr_vt_var_long}(d) shows variations in the median
velocity $V_T$ and the number of stars, $N$, calculated for
hump-creating stars. We can see  that the velocities $V_T$ (red
points) lie in the range from 220 to 245 km s$^{-1}$ except for one
time moment ($V_T=193$ km s$^{-1}$) and are shifted towards higher
values compared to $V_T$ obtained for all stars lying in the outlined
segment. The parameters of the $V_T$-velocity oscillations calculated
for hump-creating stars  are  follows: $P=1.9\pm0.1$ Myr,
$\overline{V_T}=229.7\pm0.7$ km s$^{-1}$, $A=4.48\pm1.37$ km s$^{-1}$
and $\varphi=295\pm11^\circ$. Note that the average velocity $V_T$,
calculated for  hump-creating stars ($\overline{V_T}=229.7$ km
s$^{-1}$) is larger by 11 km s$^{-1}$ than that obtained for all
stars lying in the outlined segment ($\overline{V_T}=218.5$ km
s$^{-1}$).

A comparison between Fig.~\ref{fig:vr_vt_var_long}(c) and
Fig.~\ref{fig:vr_vt_var_long}(d) shows that variations in the
velocity $V_T$ calculated for all stars in the indicated segment
(black dots) are in good agreement with oscillations in the number of
hump-creating stars $N$ (solid dark-red line) which ``pull'' the
median velocity $V_T$ towards higher values. Therefore, an increase
in $N$ causes a shift in the velocity $V_T$ towards positive values.

The influence of hump-creating  stars on the positions of the maxima
and minima of the velocity $V_T$ can also be  estimated through the
ratio of the amplitude $A$ to its error $\varepsilon_A$. The
amplitude $A$ calculated for all stars lying in the outlined segment
is $A=1.24\pm0.14$ km s$^{-1}$ while its value computed without
hump-creating stars but at the fixed values of the period and phase,
i.e. at $P=1.9\pm0.1$ Gyr  and $\varphi=329^\circ$, is only
$A=-0.28\pm0.20$ km s$^{-1}$. Thus, the ratio $|A/\varepsilon_A|$
drops from 8.9 to 1.4 after the exclusion of hump-creating stars.

Therefore, it is hump-creating stars that cause the variations in the
radial and azimuthal velocities with a period of $P=2.0\pm0.1$ Gyr.

\section{9. Distribution of model stars over the period  $P$}

\begin{figure*}
\includegraphics[width=0.8\textwidth]{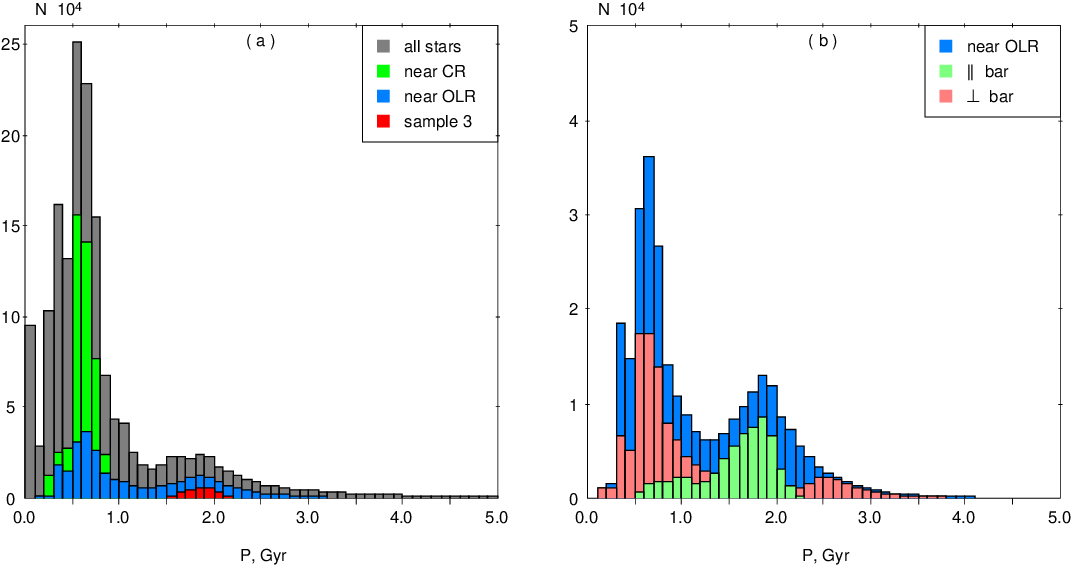}
\vspace{-0mm}     \centering \caption{Distribution of model stars
over the period $P$ of variations  in the angular momentum and
energy. (a) All  stars lying in the model disk (gray columns); stars
whose orbits lie both inside and outside the CR (green columns);
stars with orbits lying both inside and outside the OLR (blue
columns); stars creating the humps (sample 3, red columns). We can
see two maxima in the distribution of all stars over the period $P$
located at $P=0.6$ and  $P=1.9$ Gyr. Stars whose orbits lie both
inside and outside the CR (green columns) concentrate to the first
maximum. The distribution of stars with orbits lying both inside and
outside the OLR (blue columns) also has  two maxima. This
distribution in an enlarged scale is shown in the right frame. (b)
Among stars whose orbits lie both inside and outside the OLR (blue
columns), we identified orbits with the order of symmetry $m=2$
oriented perpendicular to the bar, $75\le\theta_{00}<105^\circ$
(light red columns), and parallel to the bar,
$0\le\theta_{00}<15^\circ \cup 165\le\theta_{00} < 180^\circ$
(light-green columns). The first subset of orbits (light-red columns)
has a maximum at $P=0.6$ Gyr while the maximum of the second subset
(light-green columns) corresponds to $P=1.9$ Gyr.}
\label{fig:distrib_period}
\end{figure*}

Figure \ref{fig:distrib_period} shows the distribution of stars of
the model disk over  $P$, where $P$ is the period of variations in
the angular momentum and total energy. Figure
\ref{fig:distrib_period}(a) shows the distributions built for several
samples: all stars lying in the model disk; stars whose orbits lie
both inside and outside the CR; stars with orbits lying both inside
and outside the OLR; and stars creating the humps. We can see two
maxima in the distribution of all stars over  the period $P$ located
at $P=0.6$  and  1.9 Gyr. Stars whose orbits lie both inside and
outside the CR concentrate to the first maximum.  Note that the
period of long-term oscillations near the equilibrium points $L_4$
and $L_5$ equals $565\pm2$ Myr in our model \citep{melnik2023}, so
the first maximum is likely connected with so-called banana-shaped
orbits near the  points $L_4$ and $L_5$. Hump-creating stars
concentrate to the second maximum. The distribution of stars whose
orbits lie both inside and outside the OLR (289062 stars) also has
two maxima. It is shown in an enlarged scale in the right frame.

Figure \ref{fig:distrib_period}(b) shows that the distribution of
stars with orbits lying both inside and outside the OLR also has two
maxima located at $P=0.7$ and  $P=1.9$ Gyr. We identified orbits with
the order of symmetry $m=2$ which are oriented perpendicular to the
bar  ($75\le\theta_{00}<105^\circ$) and parallel to the bar
($0\le\theta_{00}<15^\circ \cup 165\le\theta_{00}<180^\circ$), where
the angle $\theta_{00}$ determines the average direction of orbit
elongation relative to the major axis of the bar during the time
period 0--3 Gyr. It is clearly seen that orbits elongated
perpendicular to the bar (light-red columns) concentrate to the
period $P=0.7$ while orbits elongated along the bar (light-green
columns) -- to $P=1.9$ Gyr. Note that these subsets of orbits support
the outer rings $R_1$ and $R_2$, respectively.

\section{10. Age of the Galactic bar}

\label{age}

\begin{figure*}
\includegraphics[width=0.8\textwidth]{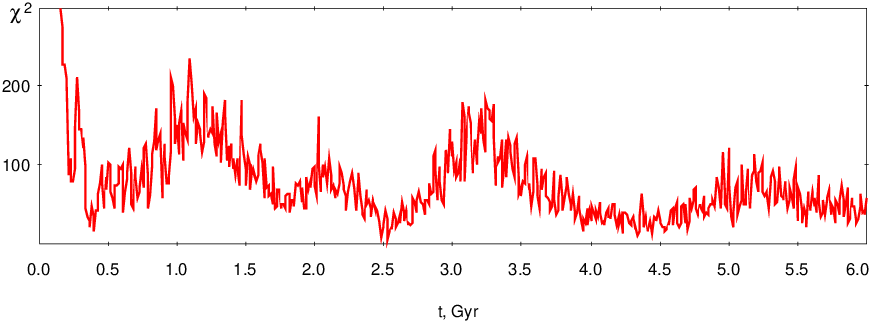}
\vspace{-0mm}     \centering \caption{Comparison between the observed
distribution of the median velocity $V_R$ derived from the {\it Gaia}
DR3 data at the distance interval $R=6$--9 kpc  and the model
distribution  at different time moments. It is  seen that the
function $\chi^2$ has two minima at the time interval 0.5--6.0 Gyr
corresponding to the  moments $t=2.5\pm0.3$ and $4.5\pm0.5$ Gyr. }
\label{fig:chi2}
\end{figure*}

We compared the  observed distribution of the median velocity $V_R$
derived from the {\it Gaia} DR3 data at the distance interval
$R=6$--9 kpc  with the similar model distribution  at different time
moments and calculated statistics $\chi^2$(t). Figure~\ref{fig:chi2}
shows the function $\chi^2$ at the time period 0--6 Gyr. The
uncertainty in the median velocity $V_R$ calculated in bins was
adopted to  be $\varepsilon_{VR}=0.6$ km s$^{-1}$ (see Section~3). We
can see that the function $\chi^2_R$ has two minima at the time
interval 0.5--6.0 Gyr corresponding to the moments $t=2.5\pm0.3$ and
$4.5\pm0.5$ Gyr. These minima arise due to the disappearance of the
humps on the model profiles of the $V_R$-velocity distributions at
the corresponding time periods (Fig.~\ref{fig:_vr_profiles}). We do
not consider the minimum at $t=0.4$ Gyr, because the bar has not
reached its full power by this moment.

The presence or absence of the humps on the model profiles of the
$V_R$-velocity distribution  can serve as an indicator of
model-observation fit. The observed profile of the $V_R$-velocity
distribution derived from {\it Gaia} DR3 data has no hump at the
distance interval $R=6$--7 kpc (Fig.~\ref{fig:_vr_profiles}). On the
other side, the model profiles have the humps at the time periods
0.6--1.8, 3.0--3.8, 5.0--5.8 Gyr. Thus, the best agreement between
the model and observations at the time interval 0.5--6.0 Gyr
corresponds to the time moments: $t=2.5\pm0.3$ and $4.5\pm0.5$ Gyr.

From what time moment should we count the age of the bar? In our
simulation, there are two critical moments: the start of simulation
and the moment when the bar reaches its full power. Real disk
galaxies almost always have at least a small oval perturbation in the
center, which can exist for a long time. So we will count the age of
the bar from the moment it reaches its full power. Since the
bar-growth time in our model is $T_g=0.45$ Gyr, the age of the bar
should be shifted by $\sim 0.5$ Gyr towards smaller values with
respect to the time moments corresponding to the best agreement
between the model and observations. Thus, the age of the Galactic bar
counted from the moment of its  reaching full strength should be
close to one of two values: $2.0\pm0.3$ or $4.0\pm0.5$ Gyr.

\section{11. Conclusions}

We studied the model of the Galaxy with a bar which reproduces well
the distributions of the observed radial, $V_R$, and azimuthal,
$V_T$, velocities of stars along the Galactocentric distance $R$
derived from {\it Gaia} DR3 data. For building the observed profiles
of the velocity distributions, we used stars lying near the Galactic
plane, $|z|<200$ pc, and in a narrow sector of the azimuthal angles,
$|\theta|<15^\circ$. The median velocities $V_R$ and $V_T$ were
calculated in $\Delta R=250$-pc wide bins. The best agreement between
the model and observed velocity profiles corresponds to the angular
velocity of the bar $\Omega_b=55\pm3$ km s$^{-1}$ kpc$^{-1}$ and the
position angle of the bar $\theta_b=45\pm15^\circ$. In our model the
radius of the OLR  and the solar Galactocentric radius are located at
the distances of $R_{OLR}=7.0$ and $R_0=7.5$ kpc, respectively
\citep{melnik2023}.

We found that the model profiles of the $V_R$-velocity distribution
demonstrate a periodic increase in the velocity $V_R$ at the distance
interval 6--7 kpc (Fig.~\ref{fig:_vr_profiles}). Maximum height of
the hump on the $V_R$-profile  equals $1.90\pm0.12$ km s$^{-1}$ and
corresponds to the distance $R=6.75$ kpc. The height of the humps
decreases with time. We studied variations in the velocity $V_R$ in
the bin  $R=6.75\pm0.125$ kpc. The average amplitude of the
$V_R$-velocity variations at the time period 0--6 Gyr is
$A=1.76\pm0.15$ km s$^{-1}$. The period of the hump formation  is
$P=2.1\pm0.1$ Gyr (Fig.~\ref{fig:_vr_vt_var}a).

The azimuthal velocity $V_T$ also shows periodic changes in  the
distance  bin $R=6.75\pm0.125$ kpc. The amplitude and period of  the
$V_T$-velocity variations are $A=1.24\pm0.14$ km s$^{-1}$ and
$P=1.9\pm0.1$ Gyr (Fig.~\ref{fig:_vr_vt_var}b).

Thus, the velocities $V_R$ and $V_T$ calculated for the segment of
the  model disk, $|\theta-\theta_0|<15^\circ$ and $R=6.75\pm0.125$
kpc, show  oscillations with a period  of $P=2.0\pm0.1$ Gyr.

Orbits trapped into librations  near the ILR and OLR demonstrate the
periodic changes in the direction of orbit elongation
\citep{weinberg1994}. We calculated the angles $\theta_{01}$,
$\theta_{02}$ and $\theta_{03}$ which determine the direction of
orbit elongation relative to the major axis of the bar during the
time periods 0--1, 1--2 and 2--3 Gyr from the start of modeling,
respectively. We found that stars whose orbits change orientation as
follows: $0<\theta_{01}<45^\circ$, $-45<\theta_{02}<0^\circ$ and
$0<\theta_{03}<45^\circ$, make significant contribution to the
formation of the humps. During the time periods 0--1 and 2--3 Gyr,
the orbits of these stars  create negative velocities $V_R$ in the
sector $|\theta-\theta_\odot|<15^\circ$ and distance range $R=6$--7
kpc while during the period of 1--2 Gyr, they change their
orientation and simply leave the region considered
(Fig.~\ref{fig:schema}). The sample of hump-creating stars includes
26308 stars which is only 9\% of all stars whose orbits lie both
inside and outside the OLR.

We studied the distribution of initial coordinates and velocities of
hump-creating stars (Fig.~\ref{fig:his_rg_2}). The maximum of the
distribution of stars over the  initial distance $R$ corresponds to
the bin $R=7.00$--7.25 kpc where maximum number of hump-creating
stars is located at $t=0$. We found the decrease in the initial
azimuthal velocity, $V_T$, with increasing initial distance, $R$. The
median initial radial velocities $V_R$ of hump-creating stars are
close to zero with the dispersion $\sim 30$ km s$^{-1}$.

A typical orbit of  hump-creating star is shown in
Figure~\ref{fig:orb_1}. Variations in the angular momentum $L$ and
total energy $E$ demonstrate  short- and long-term oscillations with
the periods of 0.13 and 2.0 Gyr, respectively. Variations in the
distance $R$ and instantaneous angular velocity $\dot{\theta}$ have
the form of beats. The angle $\theta_0$, which determines the
direction of orbit elongation with respect to the major axis of the
bar during one radial oscillation, changes in the range from $-45$ to
$45^\circ$. The angle $\theta_0$,  average distance $\overline{R}$,
average angular velocity $\overline{\dot{\theta}}$, orbital
eccentricity, and beat frequency, $w_{bt}$ (Eq.~\ref{w_bt}), change
with a period of $P=2.0\pm0.1$ Gyr.

We studied  the order of symmetry $m$ and  orientation of orbits near
the OLR. We selected orbits that lie both inside and outside the OLR
and do not cross the CR (289062 stars). Of these,  54.8\% have the
order of symmetry $m=2$; 7.7\% -- 3; 5.7\% -- 4 and 3.3\% -- 5;
28.6\% orbits have the shape close to circular. The majority of
orbits with the order of symmetry $m=2$ are stretched perpendicular
to the bar ($75\le\theta_{00}<105^\circ$) which amounts to 33.4\% of
the total sample. The fraction of orbits elongated along the bar
($0\le\theta_{00}<15^\circ \cup 165\le\theta_{00}\le 180^\circ$) is
only 16.6\% (Table~\ref{tab:statistics}, Part I).

Among  hump-creating stars (26308 stars), 88.6\% have orbits with the
order of symmetry $m=2$. The majority (71.6\%) of orbits with $m=2$
are oriented along the bar at the angle $0\le\theta_{00}<15^\circ$.
Thus, the majority of hump-creating stars support the outer ring
$R_2$. The median period calculated for  hump-creating stars is
$P=1.85$ Gyr (Table~\ref{tab:statistics}, Part II).

The fraction of librating orbits that change the direction of their
elongation within a certain range of angles,
$\theta_1<\theta_0<\theta_2$, amounts to 28\% of all orbits lying
both  inside and outside the OLR (Section~6.2).

We studied the distribution of stars whose orbits lie both inside and
outside the OLR in the plane ($E_J$, $P$). It appears that the median
period $P$ increases with decreasing $E_J$ and approaching the OLR
and then  drops sharply. Hump-creating stars demonstrate this
tendency even more clearly (Fig.~\ref{fig:jacobi}).

We explored the influence of hump-creating stars on the oscillations
of the median radial velocity $V_R$ of  stars lying in the disk
segment: $|\theta-\theta_\odot|<15^\circ$ and $R=6.75\pm0.125$ kpc
(Fig.~\ref{fig:vr_vt_var_long}). The exclusion of hump-creating stars
significantly changes the phase of the oscillations: pits appear  in
places of the humps, except for the first one, and humps arises in
places of the pits. The ratio of the  amplitude $A$ and its
uncertainty, $\varepsilon_A$, calculated for the sample without
hump-creating stars  under the fixed period and phase equals
$|A/\varepsilon_A|=1.1$ which is significantly smaller than the ratio
obtained for all stars lying in the outlined segment,
$|A/\varepsilon_A|=11.7$.

A similar test made for the azimuthal velocity $V_T$ showed that the
exclusion of  hump-creating stars under the  fixed period and phase
leads to a decrease in the ratio $|A/\varepsilon_A|$ from 8.9 to 1.4.

The distribution of  stars of the model disk over the period $P$ has
two maxima lying  at $P=0.6$ and  1.9 Gyr
(Fig.~\ref{fig:distrib_period}a). Stars whose orbits lie both inside
and outside the CR concentrate to the first maximum. Note that the
period of $P=0.6$ Gyr practically coincides with the period of
long-term oscillations around the equilibrium points $L_4$ and $L_5$,
so the first maximum is likely connected with so-called banana-shaped
orbits. The distribution of stars whose orbits lie both inside and
outside the OLR also has two maxima: at $P=0.7$ and 1.9 Gyr. Among
them, orbits elongated perpendicular and parallel to the bar
concentrate to the first and second maxima, respectively. These
orbits support the outer rings $R_1$ and $R_2$, respectively
(Fig.~\ref{fig:distrib_period}b).

A comparison between the  model and observed distributions of the
velocity $V_R$ showed that the function $\chi^2$ has two minima
corresponding to the time moments $t=2.5\pm0.3$ and $4.5\pm0.5$ Gyr
(Fig.~\ref{fig:chi2}) which arise due to the disappearance of the
humps on the model profiles of the $V_R$-velocity distribution. Thus,
the age of the Galactic bar counted from the moment of its reaching
full power must lie near one of two values: $2.0\pm0.3$ or
$4.0\pm0.5$ Gyr.

\section*{Acknowledgements}

{\footnotesize  We  thank the anonymous referees  for useful remarks
and interesting discussion. This work has made use of data from the
European Space Agency (ESA) mission {\it Gaia}
(\verb"https://www.cosmos.esa.int/gaia"), processed by the {\it Gaia}
Data Processing and Analysis Consortium (DPAC, \verb"https:"
\verb"//www.cosmos.esa.int/web/gaia/dpac/consortium"). Funding for
the DPAC has been provided by national institutions, in particular
the institutions participating in the {\it Gaia} Multilateral
Agreement. E.N. Podzolkova is a scholarship holder of the Foundation
for the Advancement of Theoretical Physics and Mathematics "BASIS"
(Grant No. 21-2-2-44-1).}

\end{document}